\documentclass[12pt]{article}
\usepackage[T1]{fontenc}
\usepackage[utf8]{inputenc}
\usepackage{authblk}

\title{\LARGE Statistical inference for association studies in the presence of binary outcome misclassification}

\author[]{Kimberly A. Hochstedler Webb}
\author[]{Martin T. Wells}
\affil[]{Department of Statistics and Data Science, Cornell University \\ Ithaca, NY}
\affil[]{kah343@cornell.edu}

\usepackage[super,sort&compress]{natbib}
\setcitestyle{comma,numbers,super,open={},close={}} 
\usepackage{graphicx}
\usepackage[paperwidth=8.5in,paperheight=11.0in,top=1in, bottom=1in, left=1in, right=1in,lines=25]{geometry}
\usepackage[title,toc,titletoc,page]{appendix}

\usepackage{siunitx}
\usepackage{appendix}
\usepackage{mathtools}
\usepackage{threeparttable}
\usepackage{booktabs}
\usepackage{bbm}
\usepackage{caption}
\usepackage{algpseudocode}
\usepackage{algorithm}
\usepackage{url}
\usepackage{xcolor}
\usepackage{lscape}

\usepackage{amsmath}
\usepackage{blkarray}

\usepackage{setspace}
\DeclareMathOperator{\logit}{logit}
\usepackage{stix}

\begin{document}

\maketitle

\begin{abstract}
In biomedical and public health association studies, binary outcome variables may be subject to misclassification, resulting in substantial bias in effect estimates. The feasibility of addressing binary outcome misclassification in regression models is often hindered by model identifiability issues. In this paper, we characterize the identifiability problems in this class of models as a specific case of ``label switching'' and leverage a pattern in the resulting parameter estimates to solve the permutation invariance of the complete data log-likelihood. Our proposed algorithm in binary outcome misclassification models \textit{does not require gold standard labels} and relies only on the assumption that the sum of the sensitivity and specificity exceeds $1$. A label switching correction is applied within estimation methods to recover unbiased effect estimates and to estimate misclassification rates. Open source software is provided to implement the proposed methods. We give a detailed simulation study for our proposed methodology and apply these methods to data from the 2020 Medical Expenditure Panel Survey (MEPS).

\end{abstract}

\newpage


\section{Introduction}
We consider regression models where a binary outcome variable is potentially misclassified. Misclassified binary outcomes are common in biomedical and public health association studies. For example, misclassification may occur in when a diagnostic test does not have perfect sensitivity or specificity. \cite{manski2021estimating, ge2023enhanced, trangucci2022identified} Misclassification can also be present in survey data, where individuals may falsely recall disease status on a self-report item. \cite{althubaiti2016information} More recently, medical studies may rely on computer algorithms to extract patient disease status from electronic medical records, but such algorithms do not perfectly capture true disease states, even when combined with record review from subject-area experts. \citep{sinnott2014improving} If analysts ignore this potential misclassification in response variables, resulting parameter estimates tend to be biased,\citep{khan2020introduction} particularly in the case of covariate-related misclassification. \citep{beesley2020statistical, zhang2020genetic} Here, we define covariate-related misclassification as the case when the probability of misclassification of an outcome variable depends on subject-level values of one or more predictors.\citep{beesley2020statistical}

Despite the known impact of covariate-related outcome misclassification, previous work on recovering unbiased association parameters is limited. Neuhaus provides general expressions for bias in the presence of covariate-related misclassification.\cite{neuhaus1999bias} Lyles et al. and Lyles and Lin extend this work, but require a validation sample or known sensitivity and specificity values, respectively, to identify association parameters.\cite{lyles2011validation, lyles2010sensitivity} Methods have also been developed to account for misclassified binary outcomes \textit{and} error-prone covariates through two-phase designs. While these methods handle cases of covariate-dependent outcome misclassification, they require an audit of a sample of the data, which in turn requires the availability of gold-standard metrics or validation procedures.\cite{lotspeich2022efficient, tang2015binary} Validation samples may not be feasible if the potentially misclassified data were obtained through a national database or through a third-party black-box algorithm. Beesley and Mukherjee address the problem of covariate-related outcome misclassification through a novel likelihood-based bias correction strategy without such a validation sample, but make the strong assumption that the binary outcome is measured with perfect specificity.\cite{beesley2020statistical} A perfect specificity assumption also underpins existing sensitivity analysis frameworks for evaluating the predictive bias of classifiers in the presence of outcome misclassification. \citep{fogliato2020fairness} Zhang and Yi develop methods to correct bias in association parameters in the context of covariate-related misclassification, but they only address the problem for mixed continuous and binary bivariate outcomes.\cite{zhang2020genetic} Our methods consider scenarios where validation data is not available, but the binary outcome is subject to covariate-dependent, bidirectional misclassification. The proposed methods also stand in contrast to those developed for error-prone categorical covariates, including the mean score method and other likelihood-based methods. \cite{han2021two, braun2017propensity}

Numerous researchers assume that sensitivity and specificity may be considered constant. \citep{magder1997logistic, daniel2003binomial, trangucci2022identified, carroll2006measurement, stamey2004parameter, xia2018bayesian, rekaya2016analysis} In such cases, misclassification rates are often either assumed to be known or estimated via validation data. \citep{magder1997logistic, trangucci2022identified, carroll2006measurement} In the event that an outcome measure has previously studied sensitivity and specificity rates, as is common for commercially available diagnostic tests, EM algorithm methods exist to incorporate that information into the fitting of logistic regression models, resulting in unbiased estimates of odds ratios. \citep{magder1997logistic} For cases where internal validation data are present, Carroll and colleagues develop general expressions for likelihood functions that take imperfect outcome measurement into account.\cite{carroll2006measurement} These methods rely on the availability of gold standard measures, which are not feasible in numerous biomedical and public health settings. \citep{Faraone1994measuring, oneill2015measuring} In contrast, fully Bayesian methods may be used to estimate constant misclassification rates in the absence of a gold standard, but such methods rely on strict and difficult prior elicitation strategies or an additional assumption of equal sensitivity and specificity. \citep{daniel2003binomial, rekaya2016analysis, rekaya2001threshold}



In addition to association studies, the impact of outcome misclassification has also been considered for estimates of  prevalence. Econometricians have used partial identifiability to address misclassification and estimate population rates of SARS CoV-2 infection,\citep{ziegler2020binary, manski2021estimating} but these methods do not readily extend to association studies. Misclassification can also impact variance estimation in prevalence studies. Ge et al. derive an estimable variance component that is induced by misclassification in certain testing applications, but their methods also rely on known sensitivity and specificity rates.\cite{ge2023enhanced} 

In the machine learning literature, the problem of outcome misclassification, or ``noisy labels'' generated from imperfect classification algorithms, is typically handled with noise-robust methods or data cleaning strategies. \citep{frenay2013classification} Other methods approach misclassification from a fairness lens, where an optimization approach is proposed to ``flip'' predicted labels after estimation to alleviate systematic bias without sacrificing the value of ``merit'' within the classification scheme. \citep{bandi2021price} When imperfect classification algorithms are used to obtain disease states, Sinnott and colleagues demonstrate that modeling outcome probabilities, rather than outcomes obtained via thresholding, improves power and estimation accuracy.\cite{sinnott2014improving} 

In this paper, we develop new strategies to recover unbiased parameter estimates in association studies with outcome misclassification that is related to observed covariates and to the true outcome. We propose a bias correction strategy that requires minimal external information, no gold standard labels, and no limitations on the misclassification patterns (i.e. perfect sensitivity or perfect specificity) present in the data.

The feasibility of addressing outcome misclassification in association studies is often limited by model identifiability issues. \citep{lyles2011validation, duan2021global} One aspect of model identifiability that must be addressed for a binary outcome misclassification model is the phenomenon of ``label switching'' found in mixture models. Label switching describes the invariance of the likelihood under relabeling of the mixture components, resulting in multimodal likelihood functions. \citep{redner1984mixture} Several remedies have been suggested to break the permutation invariance of the likelihood in both Bayesian and frequentist mixture models. \citep{rodriguez2014label, yao2015label} One common method is to impose ordering constraints on component model parameters, such as $\pi_{1} < \pi_2$. \citep{betancourt2017identifying} Such strategies aim to remove the permutation invariance of the likelihood, but are only successful for carefully chosen constraints. \citep{stephens2000dealing} In Bayesian settings, one potential resolution of the labeling degeneracy is to use non-exchangeable prior distributions to strongly separate possible parameter sets. \citep{betancourt2017identifying} It is rare, however, that analysts have enough information to set useful ordering constraints or strongly separated prior distributions in practice. \citep{rodriguez2014label} Moreover, when prior distributions for sensitivity and specificity are misspecified, bias correction approaches tend to perform poorly. \citep{ni2019comparing, daniel2003binomial} As such, we develop a novel label switching remedy that relies only on the reasonable assumption the sum of the sensitivity and specificity for the outcome measurement instrument exceeds a value of $1$. Such an assumption is recommended in other latent variable modeling approaches to misclassification problems. \cite{duan2021global, collins2014estimation, jones2010identifiability}

We use our label switching correction procedure to develop both frequentist and Bayesian estimation methods for regression parameters in an association study. This strategy also allows us to accurately estimate the average misclassification rates in the response variable and characterize the mechanism by which misclassification occurs.

In addition to label switching, several authors have considered other characteristics of misclassification models that impact identifiability. Xia and Gustafson outlines cases of unidirectional outcome misclassification where key regression parameters are identifiable. \cite{xia2018bayesian} In many practical settings, however, they find that some covariate terms can only be weakly identified. Numerical problems related to weak identifiability are also encountered the joint estimation methods by Beesley and Mukherjee (2020). \cite{beesley2020statistical} These authors address the problem by fixing model parameters at known values. In general, however, models that include a parametric observation mechanism for misclassified outcomes are identifiable if 1) perfect sensitivity or perfect specificity is assumed and 2) at least one continuous covariate is included in the true outcome mechanism but not the observation mechanism, or vice versa. \cite{beesley2020statistical, diop2011maximum}

Conditions for global identifiability of logistic regression models with misclassified outcomes are detailed in Duan et al. (2021).\cite{duan2021global} In particular, regression parameters are globally identified if there exists a subset of the covariates with support containing at least four values and with corresponding coefficients that are not exactly equal to 0. This condition is necessary and sufficient for global identifiability. It should be noted that the conditions in Duan et al. (2021) are applied to a logistic regression model with misclassified outcomes where the sensitivity and specificity parameters are \textit{not} covariate-dependent. This is in contrast to our model, where sensitivity and specificity estimates are obtained through a linear combination of parameters, estimated from a secondary logistic regression model. Given that the overall sensitivity and specificity parameter estimates are covered through the Duan et al. (2021) global identifiability conditions, the identifiability conditions that need to be considered for the misclassification-related coefficients are those that govern identifiability of logistic regression models generally. As such, we only require a full rank predictor matrix for identification of the coefficients in the component of the model governing misclassification rates. In our simulation studies and applied data example, the use of continuous predictors in the true outcome mechanism guarantees that a subset of the covariates have at least four unique values. Moreover, the simulation study contains no 0-valued coefficients by design. In the applied example, the true outcome mechanism covariates include known risk factors for a health outcome, suggesting that no coefficients would be 0-valued. In both simulation studies and the applied example, the predictor matrices are full rank. 

In Section \ref{model}, we describe the conceptual framework for our model. In Section \ref{estimation-methods}, we propose frequentist and Bayesian estimation strategies to recover the true association of interest in the presence of potentially misclassified observed outcomes. We provide software for two of the proposed estimation methods using the R package \textit{COMBO} (COrrecting Misclassified Binary Outcomes) \cite{COMBO}. Section \ref{label-switching} describes the ``label switching'' problem in greater detail and proposes a strategy to break permutation invariance of the likelihood in a binary outcome misclassification model. Through simulation, we demonstrate the utility of these methods to reduce bias in parameter estimates when compared to analyses that place restrictions on or ignore outcome misclassification. Finally, we apply our proposed methods to investigate an applied example. In particular, we study risk factors for myocardial infarction, which is known to be misdiagnosed on the basis of gender and age. \citep{arber2006patient, maserejian2009disparities, mckinlay1996non}

\section{Model, Notation, and Conceptual Framework} \label{model}
\begin{figure}[htbp]
\begin{center}
\includegraphics[scale=0.8]{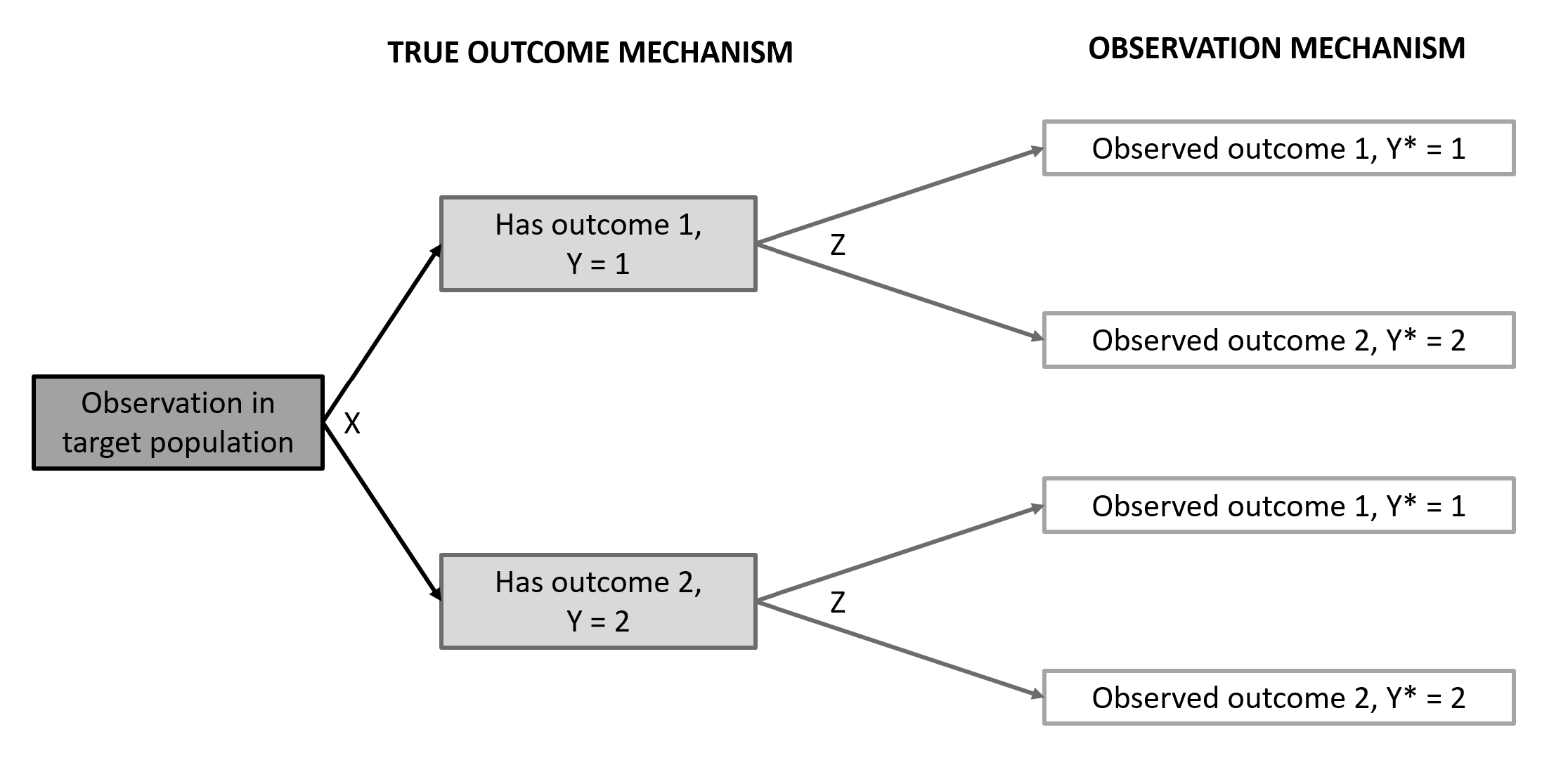}

\caption{Diagram of the assumed data structure for a binary outcome misclassification model.}\label{conceptual_framework_figure}
\end{center}
\end{figure}

Let $Y = j$ denote an observation's true outcome status, taking values $j \in \{1, 2\}$. Suppose we are interested in the relationship between the latent variable $Y$, and a matrix of predictors, $\boldsymbol{X}$, that are correctly measured. This relationship constitutes the \textit{true outcome mechanism}. Let $Y^* = k$ be a subject's observed outcome status, taking values $k \in \{1,2\}$. $Y^*$ is a potentially misclassified version of $Y$. Let $\boldsymbol{Z}$ denote a matrix of predictors related to sensitivity and specificity. The mechanism that generates the observed outcome, $Y^*$, given the true outcome, $Y$, is called the \textit{observation mechanism}. Figure \ref{conceptual_framework_figure} displays the conceptual model. The conceptual process is mathematically expressed as
\begin{equation}
\begin{aligned}
\label{eq:conceptual_framework_eq}
\text{True outcome mechanism: } &\; \logit\{ P(Y = j | \boldsymbol{X} ; \boldsymbol{\beta}) \} = \beta_{j0} + \boldsymbol{\beta_{jX} X} \\
\text{Observation mechanisms: } &\; \logit\{ P(Y^* = k | Y = 1, \boldsymbol{Z} ; \boldsymbol{\gamma}) \} = \gamma_{k10} + \boldsymbol{\gamma_{k1Z} Z}, \\
                                &\; \logit\{ P(Y^* = k | Y = 2, \boldsymbol{Z} ; \boldsymbol{\gamma}) \} = \gamma_{k20} + \boldsymbol{\gamma_{k2Z} Z}.
\end{aligned}
\end{equation}

In the true outcome mechanism, we use category $Y = 2$ as the reference category, and set all corresponding $\boldsymbol{\beta}$ parameters to 0. Similarly, $Y^* = 2$ is the reference category in the observation mechanisms, and all corresponding $\boldsymbol{\gamma}$ parameters are also set to 0. Note that the component of the \textit{observation mechanism} such that $Y = 1$ corresponds to the portion governing sensitivity. Similarly, the component of the \textit{observation mechanism} where $Y = 2$ corresponds to the specificity portion of the mechanism. Using (\ref{eq:conceptual_framework_eq}), we can express response probabilities for individual $i$'s true outcome category and for individual $i$'s observed category, conditional on the true outcome: 
\begin{flalign}
\begin{aligned}
\label{eq:response_probabilities_eq}
P(Y_i = j | \boldsymbol{X_i} ; \boldsymbol{\beta}) = &\; \; \pi_{ij} = \frac{\text{exp}\{\beta_{j0} + \boldsymbol{\beta_{jX} X_i}\}}{1 + \text{exp}\{\beta_{j0} + \boldsymbol{\beta_{jX} X_i}\}} \\
P(Y^*_i = k | Y_i = j, \boldsymbol{Z} ; \boldsymbol{\gamma}) = &\; \pi^*_{ikj} = \frac{\text{exp}\{\gamma_{kj0} + \boldsymbol{\gamma_{kjZ} Z_i}\}}{1 + \text{exp}\{\gamma_{kj0} + \boldsymbol{\gamma_{kjZ} Z_i}\}}.
\end{aligned}
\end{flalign}

These quantities can be calculated for all $N$ observations in the sample. For $j$ and $k$ both equal to the reference category, $\sum_{i = 1}^N \pi^*_{i22} = \pi^*_{22}$ measures the average specificity in the data. When $j$ and $k$ are both $1$, $\sum_{i = 1}^N \pi^*_{i11} = \pi^*_{11}$ measures the average sensitivity. Thus, (\ref{eq:response_probabilities_eq}) allows us to model sensitivity and specificity based on a set of covariates, $\boldsymbol{Z}$.

If potential misclassification in $Y^*$ is ignored by fitting a naive \textit{analysis model} $Y^* | \boldsymbol{X}$ and interpreting the results under the \textit{true outcome model} $Y | \boldsymbol{X}$,  we expect bias in $P(Y^* = j | \boldsymbol{X})$ relative to $P(Y = j | \boldsymbol{X})$. Previous work has shown that this bias is particularly severe in the case of covariate-related misclassification. \citep{beesley2020statistical} 

We define the probability of observing outcome $k$ using the model structure as
\begin{equation}
\begin{aligned}
\label{eq:p_obs_Ystar}
P(Y^* = k | \boldsymbol{X}, \boldsymbol{Z}) = \sum_{j = 1}^2 P(Y^* = k | Y = j, \boldsymbol{Z} ; \boldsymbol{\gamma}) P(Y = j | \boldsymbol{X} ; \boldsymbol{\beta}) = \sum_{j = 1}^2 \pi^*_{kj} \pi_{j}.
\end{aligned}
\end{equation}

The contribution to the likelihood by a single subject $i$ is thus $\prod_{k = 1}^2 P(Y^*_i = k | \boldsymbol{X_i}, \boldsymbol{Z_i})^{y^*_{ik}}$ where $\pi^*_{kj} = P(Y^* = k | Y = j, \boldsymbol{Z} ; \boldsymbol{\gamma})$, $\pi_{j} = P(Y = j | \boldsymbol{X} ; \boldsymbol{\beta})$, and $y^*_{ik} = \mathbbm{I}(Y^*_i = k)$, where $\mathbbm{I}(A)$ is the indicator of the set $A$. We can estimate $(\boldsymbol{\beta}, \boldsymbol{\gamma})$ using the following observed data log-likelihood for subjects $i = 1 \dots N$ as
\begin{equation}
\begin{aligned}
\label{eq:obs-log-like}
\ell_{obs}(\boldsymbol{\beta}, \boldsymbol{\gamma}; \boldsymbol{X}, \boldsymbol{Z}) = \sum_{i = 1}^N \sum_{k = 1}^2 y^*_{ik} \text{log} \{ P(Y^*_i = k | \boldsymbol{X_i}, \boldsymbol{Z_i}) \} = \sum_{i = 1}^N \sum_{k = 1}^2 y^*_{ik} \text{log} \{ \sum_{j = 1}^2 \pi^*_{ikj} \pi_{ij} \}.
\end{aligned}
\end{equation}

The observed data log-likelihood is difficult to use directly for estimation because jointly maximizing $\boldsymbol{\beta}$ and $\boldsymbol{\gamma}$ is numerically challenging for large datasets and/or datasets with even modest numbers of covariate terms. More details on these challenges are described in Section \ref{direct-obs}. 

Viewing the true outcome value $Y$ as a latent variable, we may also construct the complete data log-likelihood based on the model structure as
\begin{equation}
    \begin{aligned}
    \label{eq:complete-log-like}
    \ell_{complete}(\boldsymbol{\beta}, \boldsymbol{\gamma}; \boldsymbol{X}, \boldsymbol{Z}) &= \sum_{i = 1}^N \Bigg[ \sum_{j = 1}^2 y_{ij} \text{log} \{ P(Y_i = j | \boldsymbol{X_i}) \} + \sum_{j = 1}^2 \sum_{k = 1}^2 y_{ij} y^*_{ik} \text{log} \{ P(Y^*_i = k | Y_i = j, \boldsymbol{Z_i}) \}\Bigg] & \\
    &= \sum_{i = 1}^N \Bigg[ \sum_{j = 1}^2 y_{ij} \text{log} \{ \pi_{ij} \} + \sum_{j = 1}^2 \sum_{k = 1}^2 y_{ij} y^*_{ik} \text{log} \{ \pi^*_{ikj} \}\Bigg],
    \end{aligned}
\raisetag{12pt}\end{equation}
where $y_{ij} = \mathbbm{I}(Y_i = j)$. Without the true outcome value, $Y$, we cannot use this likelihood form directly for maximization. It is notable, however, that (\ref{eq:complete-log-like}) can be viewed as a mixture model with latent mixture components, $y_{ij}$, and covariate-dependent mixing proportions, $\pi_{ij}$.

\section{Estimation Methods} \label{estimation-methods}
In this section, we describe three estimation methods for our proposed binary outcome misclassification model. First, we describe direct optimization of the observed data log-likelihood, including resulting estimation strategies. Second, we propose jointly estimating $\boldsymbol{\beta}$ and $\boldsymbol{\gamma}$ using the Expectation-Maximization (EM) algorithm. \citep{dempster1977maximum} Next, we outline Bayesian methods for analyzing data from association studies in the presence of binary outcome misclassification. The EM algorithm and MCMC estimation strategies are available in the R package \textit{COMBO}. \cite{COMBO} All estimation strategies require corrections for label switching, described in detail in Section \ref{label-switching}.

\subsection{Direct Optimization of the Observed Data Log-likelihood} \label{direct-obs}
The observed data log-likelihood in (\ref{eq:obs-log-like}) can be directly maximized using a standard optimization package (like the \texttt{optim()} or \texttt{nlm()} functions in \texttt{R},\citep{stats2021R} for example). This method tends to be sensitive to the analysts' choice of parameter starting values, and strategies for selecting parameter starting values are discussed in Appendix \ref{optim-starting-values}. After performing direct optimization using a standard package, the procedures described in Section \ref{label-switching} are required to correct for ``label switching'' and yield final parameter estimates. Standard errors are computed by taking the square root of the diagonal entries of the inverse, negative Hessian matrix. 

\subsection{Maximization Using an EM Algorithm} \label{em}
We use the complete data log-likelihood as the starting point for the EM algorithm. Since (\ref{eq:complete-log-like}) is linear in the latent variable $y_{ij}$, we can replace $y_{ij}$ in the E-step of the EM algorithm with the quantity
\begin{equation}
\begin{aligned}
\label{eq:e-step}
w_{ij} = P(Y_i = j | Y_i^*, \boldsymbol{X}, \boldsymbol{Z}) = \sum_{k = 1}^2 \frac{y^*_{ik} \pi^*_{ikj} \pi_{ij}}{\sum_{\ell = 1}^2 \pi^*_{i k \ell} \pi_{i \ell}}.  
\end{aligned}
\end{equation}

In the M-step, we maximize the expected log-likelihood with respect to $\beta$ and $\gamma$
\begin{equation}
\begin{aligned}
\label{eq:m-step}
Q = \sum_{i = 1}^N \Bigl[ \sum_{j = 1}^2 w_{ij} \text{log} \{ \pi_{ij} \} + \sum_{j = 1}^2 \sum_{k = 1}^2 w_{ij} y^*_{ik} \text{log} \{ \pi^*_{ikj} \}\Bigr].
\end{aligned}
\end{equation}

The $Q$ function in (\ref{eq:m-step}) can be split into three separate equations for estimating $\beta$ and $\gamma$
\begin{align}
\label{eq:q-split}
Q_{\boldsymbol{\beta}} = \sum_{i = 1}^N \Bigl[ \sum_{j = 1}^2 w_{ij} \text{log} \{ \pi_{ij} \}\Bigr], \: Q_{\boldsymbol{\gamma_{k1}}} = \sum_{i = 1}^N \Bigl[\sum_{k = 1}^2 w_{i1} y^*_{ik} \text{log} \{ \pi^*_{ik1} \}\Bigr], \:
Q_{\boldsymbol{\gamma_{k2}}} = \sum_{i = 1}^N \Bigl[\sum_{k = 1}^2 w_{i2} y^*_{ik} \text{log} \{ \pi^*_{ik2} \}\Bigr].
\end{align}
$Q_{\boldsymbol{\beta}}$ contains the parameter vector $\boldsymbol{\beta}$. $Q_{\boldsymbol{\gamma_{k1}}}$ contains the $\boldsymbol{\gamma}$ parameters where the second subscript is equal to $1$, corresponding to the case where the latent variable, $Y$ is truly category $1$. $Q_{\boldsymbol{\gamma_{k2}}}$ contains the $\boldsymbol{\gamma}$ parameters where the second subscript is equal to $2$, corresponding to the case where the latent variable, $Y$ is truly category $2$. In practice, $Q_{\boldsymbol{\beta}}$ in (\ref{eq:q-split}) can be fit as a logistic regression model where the outcome is replaced by weights, $w_{ij}$. $Q_{\boldsymbol{\gamma_{k1}}}$ and $Q_{\boldsymbol{\gamma_{k2}}}$ in (\ref{eq:q-split}) are each fit as weighted logistic regression models where the outcome is $y^*_{ik}$. \citep{agresti2003categorical} After estimates for $\boldsymbol{\beta}$ and $\boldsymbol{\gamma}$ are obtained, the procedure described in Section \ref{label-switching-correction} is required to address the problem of ``label switching'' and return final parameter estimates. The covariance matrix for $\boldsymbol{\beta}$ and $\boldsymbol{\gamma}$ is obtained by inverting the expected information matrix.

\subsection{Bayesian Modeling}\label{mcmc}
Using a Bayesian approach, our proposed binary outcome misclassification model is: $Y^*_{i} | \pi^*_{i} \sim Bernoulli(\pi^*_{i})$, where $\pi^*_{i} = \sum_{j = 1}^2 \pi^*_{i1j} \pi_{ij}$ as in (\ref{eq:p_obs_Ystar}). We estimate this model using a Markov Chain Monte Carlo (MCMC) procedure. Prior distributions for the parameters are based on input from subject-matter experts on the data the model is applied to. It is recommended that prior distributions are proper and relatively flat to ensure model identifiability without strongly influencing the posterior mean estimation. For example, in cases where non-informative priors are desired, a Uniform prior with a wide range may be selected, though this selection would necessarily limit the range of resulting parameter estimates to that of the Uniform prior distribution. If an analyst has previous data suggesting a plausible estimate for a given parameter, a Normal prior distribution with a wide variance that is centered at the previous estimate may be used. In general, the choice of prior for parameters in latent class models examining misclassification is not straightforward for all settings. \cite{collins2014estimation} A more detailed description of prior elicitation strategies is provided in Appendix \ref{appendix-sim}. In the R Package, \textit{COMBO},\cite{COMBO} analysts may select between Uniform, Normal, Double Exponential, or t prior distributions, with user-specified prior parameters. Before summarizing results from MCMC, the procedure described in Section \ref{label-switching-correction} should be applied on \textit{each individual MCMC chain}, to address potential label switching within a given chain. Standard methods can be used to compute variance metrics.

\section{Label Switching} \label{label-switching}
The structure of the models described in Section \ref{model} suffers from the known problem of label switching in mixture likelihoods. Mixture likelihoods are \textit{invariant under relabeling of the mixture components}, resulting in multimodal likelihood functions. \citep{redner1984mixture} Specifically, a $J$-dimensional mixture model will have $J!$ modes in the likelihood. \citep{betancourt2017identifying} Given that our proposed model in (\ref{eq:complete-log-like}) is a mixture with $J = 2$ components labeled by the true outcome $Y \in \{1,2\}$, there are $J! = 2! = 2$ peaks in the likelihood resulting in $2$ plausible parameter sets. Thus, the multimodal nature of our proposed model likelihood means that all estimation methods described in Section \ref{estimation-methods} may converge to \textit{either} of the two plausible parameter sets. Thus, a procedure is required to obtain the parameter set of interest, given the results of the direct maximization, EM algorithm, or MCMC methods.

\subsection{Permutation Invariance of the Likelihood} \label{permutation-invariance-complete}
The invariance of the observed data log-likelihood under relabeling of mixture components is displayed in (\ref{eq:permute-invariant-obslikelihood-1}) and (\ref{eq:permute-invariant-obslikelihood-2}). In (\ref{eq:permute-invariant-obslikelihood-1}), we label $Y$ terms in the order of appearance as either $1$ or $2$. In (\ref{eq:permute-invariant-obslikelihood-2}), all terms that had $Y = 1$ are replaced with that of $Y = 2$ and all terms that had $Y = 2$ are replaced with $Y = 1$. 
\begin{equation}
\label{eq:permute-invariant-obslikelihood-1}
    \begin{aligned}
  \ell_{obs}(\boldsymbol{\beta}, \boldsymbol{\gamma}; \boldsymbol{X}, \boldsymbol{Z}) = \sum_{i = 1}^N \Bigl[ y^*_{i1} \log \{ \pi^*_{i11} \pi_{i1} + \pi^*_{i12} \pi_{i2} \} + y^*_{i2} \log \{ \pi^*_{i21} \pi_{i1} + \pi^*_{i22} \pi_{i2} \} \Bigr],
    \end{aligned}
\end{equation}
\begin{equation}
\label{eq:permute-invariant-obslikelihood-2}
    \begin{aligned}
  \ell_{obs}(\boldsymbol{\beta}, \boldsymbol{\gamma}; \boldsymbol{X}, \boldsymbol{Z}) = \sum_{i = 1}^N \Bigl[ y^*_{i1} \log \{ \pi^*_{i12} \pi_{i2} + \pi^*_{i11} \pi_{i1} \} + y^*_{i2} \log \{ \pi^*_{i22} \pi_{i2} + \pi^*_{i21} \pi_{i1} \} \Bigr]
    \end{aligned}
\end{equation}

Due to the additive structure of the observed data log-likelihood, the label change between (\ref{eq:permute-invariant-obslikelihood-1}) and (\ref{eq:permute-invariant-obslikelihood-2}) does not impact the value of the function. Thus, the observed data log-likelihood in our setting is invariant under relabeling of mixture components. 

The complete data log-likelihood is also permutation invariant under relabeling of mixture components is displayed in (\ref{eq:permute-invariant-likelihood-1}) and (\ref{eq:permute-invariant-likelihood-2}). As in (\ref{eq:permute-invariant-obslikelihood-1}) and (\ref{eq:permute-invariant-obslikelihood-2}), we label $Y$ terms in the order of appearance as either $1$ or $2$ in (\ref{eq:permute-invariant-likelihood-1}) and switch $Y = 1$ and $Y = 2$ in (\ref{eq:permute-invariant-likelihood-2}).
\begin{equation}
\label{eq:permute-invariant-likelihood-1}
    \begin{aligned}
  \ell_{complete}(\boldsymbol{\beta}, \boldsymbol{\gamma}; \boldsymbol{X}, \boldsymbol{Z}) = &\sum_{i = 1}^N \bigl[ y_{i1} \log \{ \pi_{i1} \}  + y_{i2} \log \{ \pi_{i2} \} & \\
  &\phantom{a}+ y_{i1} y^*_{i1} \log \{ \pi^*_{i11} \}  + y_{i1} y^*_{i2} \log \{ \pi^*_{i21} \}  +
   y_{i2} y^*_{i1} \log \{ \pi^*_{i12} \}  + y_{i2} y^*_{i2} \log \{ \pi^*_{i22} \} \bigr],
    \end{aligned}
\end{equation}
\begin{equation}
\label{eq:permute-invariant-likelihood-2}
    \begin{aligned}
  \ell_{complete}(\boldsymbol{\beta}, \boldsymbol{\gamma}; \boldsymbol{X}, \boldsymbol{Z}) = &\sum_{i = 1}^N \bigl[y_{i2} \log \{ \pi_{i2} \} +  y_{i1} \log \{ \pi_{i1} \} \\
  &\phantom{a} + y_{i2} y^*_{i1} \log \{ \pi^*_{i12} \}  + y_{i2} y^*_{i2} \log \{ \pi^*_{i22} \}  +
   y_{i1} y^*_{i1} \log \{ \pi^*_{i11} \}  + y_{i1} y^*_{i2} \log \{ \pi^*_{i21} \} \bigr].
    \end{aligned}
\end{equation}

Once again, the additive structure of the complete data log-likelihood means that the label change between (\ref{eq:permute-invariant-likelihood-1}) and (\ref{eq:permute-invariant-likelihood-2}) does not impact the value of the function. Thus, the complete data log-likelihood in our setting is invariant under relabeling of mixture components. Next, we will use the complete data log-likelihood to illustrate the impact of this permutation invariance on the parameter values in the function.

Suppose for each $y_{ij}$ we have a single predictor $x_i$ in the true outcome mechanism and a single predictor $z_i$ in the observation mechanisms. Substituting the parametric form of the response probabilities from (\ref{eq:response_probabilities_eq}) into the complete data log-likelihoods in (\ref{eq:permute-invariant-likelihood-1}) and (\ref{eq:permute-invariant-likelihood-2}), we can detect a pattern in the parameters associated with each likelihood mode, that is,
\begin{equation}
\label{eq:permute-invariant-param-1}
    \begin{aligned}
&\sum_{i = 1}^N \Bigl[ y_{i1}\beta_0 + y_{i1} x_i \beta_X - (y_{i1} + y_{i2}) \log \{ 1 + \exp \{ \beta_0 + x_i \beta_X \} \} \\
     &\qquad\phantom{a}+ y_{i1} y^*_{i1} \gamma_{110} + y_{i1} y^*_{i1} z_i \gamma_{11Z} - ( y^*_{i1} + y^*_{i2} ) y_{i1} \log \{1 +  \exp \{  \gamma_{110} + z_i \gamma_{11Z} \}  \} \\
     &\qquad\phantom{a}+ y_{i2} y^*_{i1} \gamma_{120} + y_{i2} y^*_{i1} z_i \gamma_{12Z} - (y^*_{i1} + y^*_{i2}) y_{i2}  \log \{ 1 +  \exp \{  \gamma_{120} + z_i \gamma_{12Z} \} \}  \Bigr] \\
 = &\sum_{i = 1}^N \Bigl[ y_{i2}(-\beta_0) + y_{i2} x_i (-\beta_X) - (y_{i1} + y_{i2}) \log \{ 1 + \exp \{ -\beta_0 + x_i (-\beta_X) \} \} \\
     &\qquad\phantom{a}+ y_{i2} y^*_{i1} \gamma_{120} + y_{i2} y^*_{i1} z_i \gamma_{12Z} - ( y^*_{i1} + y^*_{i2} ) y_{i2} \log \{1 +  \exp \{  \gamma_{120} + z_i \gamma_{12Z} \}  \} \\
     &\qquad\phantom{a}+ y_{i1} y^*_{i1} \gamma_{110} + y_{i1} y^*_{i1} z_i \gamma_{11Z} - (y^*_{i1} + y^*_{i2}) y_{i1}  \log \{ 1 +  \exp \{  \gamma_{110} + z_i \gamma_{11Z} \} \}  \Bigr].
   \end{aligned}
\end{equation}

Specifically, label switching generates the following two parameter sets:
(1) $(\beta_0, \beta_X, \gamma_{110},  \gamma_{11Z},$ 
$\gamma_{120}, \gamma_{12Z})$,
and
(2) $(-\beta_0, -\beta_X, \gamma_{120}, \gamma_{12Z}, \gamma_{110}, \gamma_{11Z})$.
Suppose the true values of each parameter were equal to real numbers $a, b, c, d, e, f$, respectively, for parameter set 1. The two parameter sets corresponding to these values would be: (1)  $a, b, c, d, e, f$, and (2) $-a, -b, e, f, c, d$. That is, the $\boldsymbol{\beta}$ parameters change signs while the $\boldsymbol{\gamma}$ parameters change $j$ subscripts between the two parameter sets. This pattern is stable for any dimension of $\boldsymbol{X}$ and $\boldsymbol{Z}$ predictor matrices. In practice, if an analyst uses frequentist estimation methods, they will recover only one of the two parameter sets. When this model is fit via Gibbs Sampling, it is typical for different chains to center at different parameter sets. Moreover, each chain is generally unable to transition \textit{between} modes within a finite running time. \citep{betancourt2017identifying} Thus, it is possible for a final posterior sample to reflect estimates from different ``label-switched'' parameter sets, making the results difficult to meaningfully summarize. \citep{stephens2000dealing}

\subsection{A Strategy to Correct Label Switching}\label{label-switching-correction}
Given that the two likelihood modes present in the proposed model are known, we now describe a method to select the appropriate parameter set. After obtaining parameter estimates, $\boldsymbol{\hat{\beta}}$ and $\boldsymbol{\hat{\gamma}}$, for the proposed misclassification model using an estimation method described in Section \ref{estimation-methods}, the appropriate parameter set, $\boldsymbol{\hat{\beta}}_{corrected}$ and $\boldsymbol{\hat{\gamma}}_{corrected}$, can be determined using the following procedure.

First, average correct classification rates should be estimated using parameter estimates $\boldsymbol{\hat{\beta}}$ and $\boldsymbol{\hat{\gamma}}$ obtained from either direct maximization of the observed data log-likelihood (Section \ref{direct-obs}), the proposed EM algorithm (Section \ref{em}), or \textit{a single chain} of the proposed MCMC procedure (Section \ref{mcmc}). These estimated averages, denoted $\hat{\pi}^*_{jj}$ are taken across all observations $i$, for both $j = 1$ and $j = 2$, as follows: 
\begin{equation}
\begin{aligned}
\label{eq:avg_pistar}
\hat{\pi}^*_{11} = \sum_{i = 1}^N \hat{\pi}^*_{i11} = \sum_{i = 1}^N \frac{\text{exp}\{\hat{\gamma}_{110} + \boldsymbol{\hat{\gamma}_{11Z} Z_i}\}}{1 + \text{exp}\{\hat{\gamma}_{110} + \boldsymbol{\hat{\gamma}_{11Z} Z_i}\}},\\
\hat{\pi}^*_{22} = \sum_{i = 1}^N \hat{\pi}^*_{i22} = \sum_{i = 1}^N \frac{1}{1 + \text{exp}\{\hat{\gamma}_{120} + \boldsymbol{\hat{\gamma}_{12Z} Z_i}\}},
\end{aligned}
\end{equation}

When $j = 1$, this quantity estimates the average sensitivity across all observations. When $j = 2$, this quantity estimates the average specificity across all observations.

Based on the label switching patterns described in Section \ref{label-switching}, the following relationship exists between the estimated sensitivity and specificity computed using $\boldsymbol{\hat{\beta}}$ and $\boldsymbol{\hat{\gamma}}$, denoted $\hat{\pi}^*_{11}$ and $\hat{\pi}^*_{22}$ respectively, and the estimated sensitivity and specificity computed using the label-switched parameter set, denoted $\hat{\pi}^{*, switch}_{11}$ and $\hat{\pi}^{*, switch}_{22}$ respectively:
\begin{equation}
\begin{aligned}
\label{eq:misclassification_label_switch}
\hat{\pi}^{*, switch}_{11} = 1 - \hat{\pi}^*_{22},\\
\hat{\pi}^{*, switch}_{22} = 1 - \hat{\pi}^*_{11}.
\end{aligned}
\end{equation}

Details on this relationship are available in Appendix \ref{appendix-label-switch}. Using this relationship, we compute $\hat{\pi}^{*, switch}_{11}$ and $\hat{\pi}^{*, switch}_{22}$, the estimated average sensitivity and specificity, respectively, under the label-switched parameter set. 

Next, we compute Youden's $J$ Statistic, a composite measure of the performance of a classifier,\cite{BERRAR2019546} for both the estimated and label-switched parameter sets, denoted $\hat{J}$ and $\hat{J}^{switch}$, respectively:
\begin{equation}
\begin{aligned}
\label{eq:J-stats}
&\hat{J} = \hat{\pi}^*_{11} + \hat{\pi}^*_{22} - 1,\\
&\hat{J}^{switch} = \hat{\pi}^{*, switch}_{11} + \hat{\pi}^{*, switch}_{22} - 1.
\end{aligned}
\end{equation}

In order to select the appropriate parameter set, $\boldsymbol{\hat{\beta}}_{corrected}$ and $\boldsymbol{\hat{\gamma}}_{corrected}$, we compare $\hat{J}$ and $\hat{J}^{switch}$. If $\hat{J} \geq \hat{J}^{switch}$, then $\boldsymbol{\hat{\beta}}_{corrected} \gets \boldsymbol{\hat{\beta}}$ and $\boldsymbol{\hat{\gamma}}_{corrected} \gets \boldsymbol{\hat{\gamma}}$. That is, the appropriate parameter set is that which was obtained using methods from Section \ref{estimation-methods} and no further action is required. Otherwise, if $\hat{J} < \hat{J}^{switch}$, then $\boldsymbol{\hat{\beta}}_{corrected} \gets -\boldsymbol{\hat{\beta}}$, $\boldsymbol{\hat{\gamma}}_{corrected, k1} \gets \boldsymbol{\hat{\gamma}}_{k2}$, and $\boldsymbol{\hat{\gamma}}_{corrected, k2} \gets \boldsymbol{\hat{\gamma}}_{k1}$. That is, the appropriate parameter set is actually the label-switched parameter set. Described in words, if the appropriate parameter set is deemed to be the label-switched parameter set, then the $\boldsymbol{\hat{\beta}}$ vector is multiplied by $-1$ and the second index for every element of $\boldsymbol{\hat{\gamma}}$ is switched (i.e. all second indices that were ``1'' become ``2'' and all second indices that were ``2'' become ``1'') in order to obtain $\boldsymbol{\hat{\beta}}_{corrected}$ and $\boldsymbol{\hat{\gamma}}_{corrected}$, respectively. This procedure is also presented as an algorithm using pseudocode in Appendix \ref{appendix-algorithm}. If $\boldsymbol{\hat{\beta}}$ and $\boldsymbol{\hat{\gamma}}$ were obtained from direct maximization of the observed data log-likelihood or from the proposed EM algorithm, then this procedure yields the final parameter estimates. If $\boldsymbol{\hat{\beta}}$ and $\boldsymbol{\hat{\gamma}}$ were obtained from a single MCMC chain, then this procedure should be independently executed on each chain. The resulting $\boldsymbol{\hat{\beta}}_{corrected}$ and $\boldsymbol{\hat{\gamma}}_{corrected}$ estimates from each chain should be combined to estimate posterior sample quantities. 

In addition to selecting the appropriate parameter set, the information matrix must also be corrected in the presence of label switching. If $\hat{J} \geq \hat{J}^{switch}$, the information matrix obtained from a given estimation procedure is the appropriate information matrix. If $\hat{J} < \hat{J}^{switch}$, then the information matrix requires an adjustment. Specifically, the rows and columns of the information matrix should be relabeled such that all rows and columns corresponding to $\boldsymbol{\hat{\gamma}}_{k2}$ terms now correspond to $\boldsymbol{\hat{\gamma}}_{k1}$ terms. Similarly, all rows and columns corresponding to $\boldsymbol{\hat{\gamma}}_{k1}$ terms now correspond $\boldsymbol{\hat{\gamma}}_{k2}$ terms. If this correction is performed, the information matrix is inverted and summarized as usual, but now the position of the variance terms from the matrix correspond to different parameters. Such a transformation is illustrated in an example in Appendix \ref{appendix-info-matrix-switch}.

The described procedure is based on the assumption that the appropriate parameter set (either the estimated or the label-switched parameter set) is that which jointly maximizes the estimated sensitivity and specificity. Equivalently, we assume that the sum of the sensitivity and the specificity must be greater than $1$ for the appropriate parameter set (see Appendix \ref{appendix-label-switch}). This assumption is required in order to differentiate between the two possible parameter sets for this model. As displayed in (\ref{eq:permute-invariant-likelihood-1}) and (\ref{eq:permute-invariant-likelihood-2}), each parameter set assigns a different category of the latent variable, $Y = 1$ or $Y = 2$, to each component of the mixture likelihood. This means that when we compute $\pi^*_{ikj} = P(Y_i^* = k | Y_i = j)$ for a given individual, $i$, and observed outcome category, $k$, it is possible that $\pi^*_{ikj}$ could refer to $\pi^*_{ik1}$ or $\pi^*_{ik2}$. In order to differentiate between these two latent variable assignments, we make the assumption that the appropriate parameter set will yield a larger value of Youden's $J$ Statistic than that of the other parameter set that maximizes likelihood. Thus, we are able to determine which latent category labeling is more plausible in our results. This assumption is reasonable because it discriminates between candidate parameter sets by favoring that which has larger additive correct classification rates. Moreover, this criteria is commonly used to eliminate label switching problems in other misclassification models that use latent class analysis methods. \cite{collins2014estimation, duan2021global, lamont2016regression, jones2010identifiability}

Such assumptions are not required in other methods like Beesley and Mukherjee (2020) and Xia and Gustafson (2018) because unidirectional misclassification is assumed. \cite{beesley2020statistical, xia2018bayesian} Thus, only one latent outcome category is present in the complete data log-likelihood and so the resulting functional form is not permutation invariant.

\section{Simulations} \label{simulations}
We present simulations for evaluating the proposed binary outcome misclassification model in terms of bias and root mean squared error (rMSE). Our proposed direct maximization, EM, and MCMC methods are compared to two versions of the method in the \textit{SAMBA} R package and to a naive logistic regression that assumes there is no measurement error in the observed outcome. In the \textit{SAMBA} methods, the \textit{observation mechanism} is assumed to have either perfect specificity or perfect sensitivity, denoted \textit{SAMBA}-Specificity and \textit{SAMBA}-Sensitivity, respectively. \cite{samba} Details on prior distribution specification for the $\boldsymbol{\beta}$ and $\boldsymbol{\gamma}$ parameters are available in Appendix \ref{appendix-sim} and Table \ref{sim-setting-table1}.

The model simulation study includes three settings: (1) small sample size and large misclassification rates, (2) large sample size and small misclassification rates, and (3) moderate sample size with perfect specificity. Details on these settings can be found in Appendix \ref{appendix-sim}. In particular, details on how $\boldsymbol{X}$ and $\boldsymbol{Z}$ were generated as well as the true values of regression coefficients $\boldsymbol{\beta}$ and $\boldsymbol{\gamma}$ are provided in Appendix \ref{appendix-sim-data-gen}.

Figures 2 and 3 present the parameter estimates for Setting 1 and Setting 2, respectively, across 500 simulated datasets. Figure \ref{fig:ps_sim_results_histogram} presents parameter estimates under Setting 3. Figure \ref{fig:ps_sim_results_histogram_zoom} truncates the range of the presented parameter estimates from Figure \ref{fig:ps_sim_results_histogram} to show more detail of the distribution of $\gamma_{120}$ and $\gamma_{12Z}$ estimates in Setting 3. Table \ref{parameter-results-table} presents mean parameter estimates and rMSE across 500 simulated datasets for each setting and estimation method. Table \ref{probability-results-table} presents the outcome probability, sensitivity, and specificity values measured from the generated data and estimated from each method.

\textbf{Setting 1:} Across all simulated datasets, the average $P(Y = 1)$ was 64.7\% and the average $P(Y^* = 1)$ was 59.1\%. The empirical correct classification rate for $Y = 1$, or the average sensitivity, was 84.7\%. The empirical correct classification rate for $Y = 2$, or the average specificity, was 87.7\%  (Table \ref{probability-results-table}). In Table \ref{parameter-results-table}, we see that the naive logistic regression and the \textit{SAMBA}-Sensitivity analyses result in substantial bias the estimate of $\beta_0$ and $\beta_X$. If perfect specificity is assumed using \textit{SAMBA}-Specificity, the bias and rMSE of the parameter estimates is moderate. For the main association parameter of interest, $\beta_X$, the relative bias of the \textit{SAMBA}-Specificity estimate reaches $20\%$. In contrast, our proposed EM algorithm and direct maximization procedure have relative bias of $5\%$ and $1\%$, respectively, for the $\beta_X$ estimate. Across all estimated parameters, the direct maximization and EM procedures perform well in Setting 1, though direct maximization generally produces lower rMSEs than the EM approach. This is especially true for estimates of the slope term of the observation mechanism for specificity, $\gamma_{12Z}$, where the EM algorithm estimate has a relative bias of over $30\%$. This finding is likely a result of the characteristics of this particular simulation setting. Namely, $P(Y = 2)$ is somewhat small, $0.353$. This fact, combined with the ``small'' sample size of $1,000$, means that the raw number of observations that were subject to imperfect specificity was quite low. In turn, this makes Frequentist estimation with the EM algorithm more challenging. Compared to the proposed EM Algorithm and direct maximization procedure, the MCMC method in Section \ref{mcmc} performs poorly in Setting 1 in terms of bias. Despite the larger bias, however, the estimates from MCMC tend to have lower variation than those from the EM algorithm, perhaps due to the influence of the Normal prior distributions used for each parameter. 


In Table \ref{probability-results-table}, we observe that the direct maximization and EM algorithm approaches produce accurate estimates of $P(Y = 1)$, with an absolute bias of $0.001$ and $0.002$, respectively. The MCMC approach underestimates $P(Y = 1)$, on average, by $0.022$ (in absolute terms). Methods that make assumptions about the misclassification mechanisms also show bias in their outcome prevalence estimates. The \textit{SAMBA}-Specificity, \textit{SAMBA}-Sensitivity, and naive approaches resulted in relative bias of $4\%$, $11\%$, and $9\%$, respectively, for their estimates of $P(Y = 1)$. All of the methods that produced estimates of sensitivity, $P(Y^* = 1 | Y = 1)$, overestimated the empirical value. The bias was greatest for MCMC and \textit{SAMBA}-Specificity, which produced estimates with a relative bias of $3\%$ and $4\%$, respectively. The proposed EM algorithm, MCMC, and direct maximization approaches all slightly underestimated the specificity, $P(Y^* = 2 | Y = 2)$. \textit{SAMBA}-Sensitivity, on the other hand, overestimated specificity by 0.084 (in absolute terms).

It should be noted that, despite the parallel nature of the \textit{SAMBA}-Sensitivity and \textit{SAMBA}-Specificity methods, the \textit{SAMBA}-Sensitivity approach had worse performance. This difference is attributable to the nature of the simulation study. In particular, $P(Y = 1)$ is considerably larger than $P(Y = 2)$ due to the selection of parameter values. This fact, combined with the relatively similar values for sensitivity and specificity, meant that more misclassification naturally occurred for subjects with a true outcome value of $1$ than for subjects with a true outcome value of $2$. Thus, the perfect sensitivity assumption in \textit{SAMBA}-Sensitivity has a greater impact on the results than the perfect specificity assumption in \textit{SAMBA}-Specificity, simply due to the nature of the study design. Under settings where $P(Y = 2) > P(Y = 1)$ and $P(Y^* = 2 | Y = 2) \approx P(Y^* = 1 | Y = 1)$, this trend reverses, as expected (data not shown).

\textbf{Setting 2:} In Setting 2, the average $P(Y = 1)$ was 64.8\% and the average $P(Y^* = 1)$ was 61.8\% across all simulated datasets (Table \ref{probability-results-table}). The sensitivity was 92.4\% and the average specificity was 94.5\%. There is substantial bias in the estimates of $\beta_0$ and $\beta_X$ when naive logistic regression and \textit{SAMBA}-Sensitivity approaches are used (Table \ref{parameter-results-table}), indicating that even relatively mild misclassification can severely impact parameter estimation, despite a large sample size. When perfect specificity assumed in \textit{SAMBA}-Specificity the bias and rMSE of the $\boldsymbol{\beta}$ estimates are improved compared to naive approaches, but are still higher than the direct maximization, EM algorithm and MCMC procedures that account for both imperfect sensitivity and imperfect specificity (Table \ref{parameter-results-table}). The proposed direct maximization and EM algorithm approaches also perform well in terms of bias and rMSE for the $\boldsymbol{\gamma}$ parameters in Setting 2. The MCMC approach, however, produces estimates with larger bias and rMSE for all $\boldsymbol{\gamma}$ parameters compared to the other two novel procedures. We attribute this poorer performance to the prior specification. We suspect that the prior mean, which was not centered at the true parameter values, may have influenced final explanation in a detrimental way in this setting.

In Table \ref{probability-results-table}, we see that all proposed methods recover the true outcome probabilities with virtually no bias. In contrast, the \textit{SAMBA}-Specificity approach overestimates $P(Y = 1)$ by $0.013$, on average, while the \textit{SAMBA}-Sensitivity and naive approaches underestimate $P(Y = 1)$ by $0.034$ and $0.030$, on average, respectively. All tested methods slightly overestimate sensitivity in Setting 2. The relative bias of $P(Y^* = 1 | Y = 1)$ estimates are between $0.6\%$, and $0.7\%$ for the direct maximization, EM algorithm, and MCMC approaches. Relative bias for the estimate of $P(Y^* = 1 | Y = 1)$ is slightly higher for \textit{SAMBA}-Specificity, at $2\%$. Our proposed approaches also recover reasonable estimates of the specificity, $P(Y^* = 2 | Y = 2)$. The relative bias of estimates of the specificity from the direct maximization, EM algorithm, and MCMC approaches is between $1\%$ and $2\%$. \textit{SAMBA}-Sensitivity overestimates the empirical specificity by 0.043 (in absolute terms). In Setting 2, \textit{SAMBA}-Specificity again shows better performance than \textit{SAMBA}-Sensitivity due to the particular simulation setting characteristics (described with Setting 1). 

\textbf{Setting 3:} In Setting 3, the average $P(Y = 1)$ was 64.7\% and the average $P(Y^* = 1)$ was 54.8\% (Table \ref{probability-results-table}). Per the simulation design, there was no misclassification for $Y = 2$ (i.e. perfect specificity). The average sensitivity was 84.6\%. By designing a simulation setting where a perfect specificity assumption is appropriate, we are testing the performance of our proposed methods in a case of model misspecification. In our model specification, exactly perfect specificity corresponds to a case where $\gamma_{120}$ and $\gamma_{12Z}$ parameters are valued at negative infinity. In practice, setting these parameters as $\gamma_{120} = -5$ and $\gamma_{12Z} = -5$ produces the desired perfect specificity effect empirically. The fact remains though, that procedures for estimating $\gamma_{120}$ and $\gamma_{12Z}$ in cases of perfect or near-perfect specificity may be more challenging than other scenarios. This simulation allows for investigation of this model misspecification scenario.

In Table \ref{parameter-results-table}, we see that the naive approach and the \textit{SAMBA}-Sensitivity approach both yield highly biased $\boldsymbol{\beta}$ parameter estimates. This behavior can be attributed to the fact that the naive approach does not account for misclassification and \textit{SAMBA}-Sensitivity only allows for imperfect specificity; both of these assumptions are far from appropriate in Setting 3. As expected, the \textit{SAMBA}-Sensitivity method also yields highly biased estimates for $\gamma_{120}$ and $\gamma_{12Z}$, as these parameters govern the specificity component of the observation mechanism, which we know to be error-free in this context.

The framework of Setting 3 matches the \textit{SAMBA}-Specificity method developed by Beesley and Mukherjee,\cite{beesley2020statistical} and it is unsurprising that low bias and rMSE is observed across all  \textit{SAMBA}-Specificity parameter estimates. Despite making no assumptions on misclassification direction in a setting with perfect specificity, the parameter estimates for the proposed EM algorithm and direct maximization approaches have low bias for all $\boldsymbol{\beta}$ terms and for $\gamma_{110}$ and $\gamma_{11Z}$. While the bias for the $\boldsymbol{\beta}$, $\gamma_{110}$, $\gamma_{11Z}$ estimates from the proposed MCMC method are slightly higher than that observed from direct maximization or the EM algorithm, it still only reaches a relative magnitude of between $4\%$ and $11\%$. Moreover, the spread of MCMC parameter estimates is smaller than that of estimates obtained from the other two novel methods. Substantial bias is observed in the $\gamma_{120}$ and $\gamma_{12Z}$ estimates obtained from the direct maximization, EM algorithm, and MCMC approaches. This bias is largely the result of extreme parameter estimates in some simulated datasets, as observed in Figure \ref{fig:ps_sim_results_histogram}. Despite the extreme variation in the $\gamma_{120}$ and $\gamma_{12Z}$ estimates across simulations, it should be noted that the remaining parameters are estimated with low bias and with rMSE near that of the \textit{SAMBA}-Specificity estimates (Table \ref{parameter-results-table}). Thus, despite our model's incorrect assumption of imperfect specificity, parameter estimates for the $\beta_0$, $\beta_X$, $\gamma_{110}$, and $\gamma_{11Z}$ coefficients are still highly accurate.

In Table \ref{probability-results-table}, the EM, direct maximization, and \textit{SAMBA}-Specificity methods return accurate estimates of $P(Y = 1)$ and $P(Y = 2)$. In contrast, the MCMC approach yields a slight underestimate of $P(Y = 1)$ (relative bias, $3\%$) and the naive analysis yields a severe underestimate of $P(Y = 1)$ (relative bias, $15\%$). The direct maximization, EM algorithm, MCMC, and \textit{SAMBA}-Specificity approaches all perform well for estimates of sensitivity, $P(Y^* = 1 | Y = 1)$, with absolute bias between 0.009 and 0.012. Importantly, the direct maximization, EM algorithm, and \textit{SAMBA}-Sensitivity approaches yield accurate specificity estimates. The MCMC approach yields a slightly less accurate specificity estimate, with absolute bias of 0.058. It should be noted that the novel direct maximization and EM algorithm procedures recovered near-perfect specificity estimates, despite high bias in the $\gamma_{120}$ and $\gamma_{12Z}$ terms. Moreover, the direct maximization and EM algorithm approaches yielded accurate parameter estimates for all other terms ($\beta_0$, $\beta_X$, $\gamma_{110}$, and $\gamma_{11Z}$). This suggests that the proposed methods are robust to this form of model misspecification. In particular, in the event that there is perfect specificity present in the data, the proposed model can recover a $P(Y^* = 2 | Y = 2)$ estimate near 1. With such a high specificity estimate, analysts can detect that the perfect specificity assumption is appropriate, and switch to analyzing the data using the \textit{SAMBA}-Specificity approach.

\section{Applied Example} \label{example} 
In this section, we perform a case study using data from the 2020 Medical Expenditure Panel Survey (MEPS), a sequence of surveys on cost and use of health care and health insurance status in the United States. \cite{AHRQ} We are primarily interested in risk factors associated with myocardial infarction (MI). This association, however, is potentially impacted by misclassification in self-reported history of MI. In particular, it is possible for the symptoms of other ailments to be attributed to MI and for the chest pain of MI to be attributed to a different diagnosis. These types of misdiagnoses are known to be more common in women than men and among younger patients. \citep{arber2006patient, maserejian2009disparities, mckinlay1996non} Thus, we expect misclassification of self-reported history of MI to be associated with patient gender and age. Our overall goal is to assess the severity of this misclassification, and to understand the impact of misclassification on the association between risk factors and MI. The MEPS 2020 survey was used for this analysis because, at the time of writing, this was the most recently collected MEPS data. In addition, the disruption of the COVID-19 pandemic in the 2020 year may have contributed additional measurement error to medical diagnoses in the dataset. \citep{AHRQ}

Our response variable of interest is whether or not survey respondents reported any history of MI. We assessed the association of MI status with risk factors including age, smoking status, and exercise habits. The age variable was centered and scaled before use in the model. Smoking status and exercise variables were coded as binary indicators. The use of a continuous age variable ensures that we have at least four unique covariate values in this setting. In addition, the true outcome mechanism covariates include known risk factors for MI, suggesting that no $\boldsymbol{\beta}$ coefficients would be 0-valued. Finally, the $\boldsymbol{Z}$ matrix is full rank. Thus, we have satisfied the conditions for identification of a logistic regression model with misclassified outcomes and covariate-dependent misclassification rates.

For the analysis, we included data only from participants age 18-85. For each family unit recorded in the dataset, we only keep the responses of the reference person (defined as the household who owns or rents the home and is over 16 years of age or the head of the household). \citep{AHRQ} After excluding all of the records with missing values in the response and covariates, we had a total of 12731 observations on our analysis. In this dataset, 640 (5\%) reported a history of MI.

We estimate model parameters using our proposed direct maximization, EM algorithm, and MCMC approaches with the label switching correction procedure from Section \ref{estimation-methods}. For the MCMC approach, we selected prior distributions based on the principles described in Appendix \ref{appendix-sim}. All $\boldsymbol{\beta}$ parameters had \textit{Normal} priors, centered at estimates from the naive \textit{analysis model} with a standard deviation of 10. The prior distributions for the $\gamma_{11,gender}$, $\gamma_{11,age}$, $\gamma_{12,gender}$, $\gamma_{12,age}$ parameters were all \textit{Normal}$(0, 10)$, suggesting a prior belief that patient gender and age are unrelated to MI misdiagnosis. For $\gamma_{110}$, we selected a \textit{Normal}$(1, 10)$ prior distribution. For $\gamma_{120}$, we selected a \textit{Normal}$(-2, 10)$ prior distribution. These selections result in prior sensitivity and specificity values of 0.731 and 0.881, respectively. These prior sensitivity and specificity values were selected because previous literature indicated that MI diagnosis tends to have higher sensitivity than specificity.\cite{park2018sensitivity} We compare these novel methods to the \textit{SAMBA}-Specificity, \textit{SAMBA}-Sensitivity, and naive approaches that assume perfect specificity, perfect sensitivity, or no misclassification in the outcome measurement instrument, respectively. Parameter estimates and standard errors are reported in Table \ref{applied-results-table}.

The $\beta_0$, $\beta_{smoke}$, and $\beta_{exercise}$ estimates in the \textit{SAMBA}-Specificity and naive analysis methods are all attenuated compared to the proposed methods. In contrast, the $\beta_{age}$ estimate is smaller in the methods that allow for imperfect sensitivity and specificity, compared to that from the \textit{SAMBA}-Specificity and naive analysis methods. Using any of the proposed estimation methods, we find that smoking, not exercising regularly, and increased age are all associated with true incidence of MI in the sample. In the sensitivity component of the \textit{observation mechanism}, estimates from the direct maximization and EM algorithm approaches suggest that, given that MI has occurred, individuals who are older and/or women are less likely to report having been diagnosed with MI (Direct Max.: $\gamma_{11,gender} = -1.768$ and $\gamma_{11,age} = -0.199$; EM: $\gamma_{11,gender} = -1.766$ and $\gamma_{11,age} = -0.198$). The MCMC approach did not find a significant association between patient gender and patient age in the sensitivity mechanism (MCMC: $\gamma_{11,gender} = -0.142$ with \textit{SE} = 0.315 and $\gamma_{11,age} = 0.107$ with \textit{SE} = 0.243). Given that MI has not occurred, estimates from the direct maximization and EM algorithm approaches suggest that women are still less likely to report an MI diagnosis, but older individuals are more likely to report having been diagnosed with MI (Direct Max.: $\gamma_{12,gender} = -0.818$ and $\gamma_{12,age} = 0.083$; EM: $\gamma_{12,gender} = -0.818$ and $\gamma_{12,age} = 0.084$). The same trends are detected in the MCMC analysis (MCMC: $\gamma_{12,gender} = -0.975$ and $\gamma_{12,age} = 0.072$). Perfect sensitivity and perfect specificity assumptions reveal highly variable estimates of associated $\boldsymbol{\gamma}$ parameters using the \textit{SAMBA}-Specificity and \textit{SAMBA}-Sensitivity methods. Given that our simulation studies suggested that the EM algorithm and direct maximization approaches yielded more accurate parameter estimates in most scenarios, and that the EM algorithm and direct maximization parameter estimates were nearly equal, we will only interpret the results under the EM algorithm going forward. 

Using the EM $\boldsymbol{\beta}$ terms, we estimate the true prevalence of MI to be $5\%$. We also use the $\boldsymbol{\gamma}$ parameter estimates from the EM algorithm approach to estimate the sensitivity and specificity of the self-reported MI measure among men and women in the sample. Among men, the sensitivity is estimated at 76.3\% and the specificity is estimated at 94.4\%. Among women, sensitivity is much lower, at only 59.1\%, and specificity is estimated at 97.1\%. These findings are in line with previous literature which suggests that many instances of MI go misdiagnosed or undiagnosed,\citep{turkbey2015prevalence} and this problem is more pronounced for women in comparison to men. \citep{arber2006patient, maserejian2009disparities}

\section{Discussion}
Association studies that use misclassified outcome variables are susceptible to bias in effect estimates. \citep{beesley2020statistical, khan2020introduction} In this paper, we presented methods to recover unbiased regression parameters when a binary outcome variable is subject to misclassification and provide software in the \texttt{R} package \textit{COMBO}. \cite{COMBO} These methods are based a novel label switching correction procedure, which is required to overcome model permutation invariance problems inherent in the likelihood structure. We show that our methods are able to recover parameter estimates in cases with varying sample sizes and misclassification rates, compared to methods that assume perfect sensitivity, perfect specificity, or no misclassification in the outcome measure. To show the utility of our method in real-world problems, we apply our methods to data from the 2020 Medical Expenditure Panel Survey (MEPS) and estimate that sensitivity and specificity rates for MI diagnosis differ by patient gender and age.

Our methods are attractive because they require little external information, can be implemented without repeated measures or double-sampled outcomes, and do not require gold standard labels. Our findings also constitute a generalization of the work of Beesley and Mukherjee (2020). \cite{beesley2020statistical} Specifically, we no longer require the assumption of perfect specificity, but can still handle cases where such an assumption is valid. Moreover, our simulation results showcase the potential risk of assuming the nature of the misclassification (i.e. perfect sensitivity or perfect specificity) in an analysis. Depending on response probabilities of the latent outcome, inappropriately deploying \textit{SAMBA}-Sensitivity or \textit{SAMBA}-Specificity can result in parameter estimates that are on par with or worse than those from methods that do not account for misclassification at all (see Setting 1 and Setting 2 results in Section \ref{simulations}). Using our methods allows analysts to avoid \textit{a priori} assumptions of perfect specificity or perfect sensitivity, thus avoiding this bias in the event that the assumption is incorrect.

Further generalizations of our methods are still possible in future work. In particular, researchers may consider the more difficult case of a potentially misclassified outcome that can take on three category labels. In addition, our work assumes perfect measurement of predictors in the model. This assumption is unlikely to always hold in practice. For example, self-reported smoking status and exercise may not be reliable measures in the MEPS MI example in Section \ref{example}. Thus, a valuable extension of this work could examine cases of misclassification in both the outcome variable and the predictors simultaneously. Existing likelihood-based and mean score methods may shed light on approaches to handle exposure misclassification in such settings, though these approaches require validation samples. \cite{braun2017propensity, han2021two, lotspeich2022efficient, tang2015binary} Additional research directions can also explore how covariate-related misclassification can impact variables other than outcomes.

\subsection*{Acknowledgements}
The authors would like to thank Yu Cheng for her advice and expertise on the label switching correction procedure, and the inspiration to use Youden's $J$ Statistic. In addition, we thank the anonymous reviewers for their comments and suggestions, which led to substantive improvements in the paper. 

Funding support for KHW was provided by the LinkedIn and Cornell Ann S. Bowers College of Computing and Information Science strategic partnership PhD Award. MW was supported by NIH awards U19AI111143-07 and 1P01-AI159402.

\subsection*{Data Accessibility}
The data used in the applied example are freely available for download from the Medical Expenditures Panel Survey.

\subsection*{Code Availability}
Open source software is provided to implement the proposed methods. The R package can be downloaded at https://cran.r-project.org/web/packages/COMBO/

\newpage

\bibliographystyle{ama_with_first_initial_only}
\bibliography{references}

\newpage
\section{Tables}

\begin{landscape}

\begin{table}[H]\footnotesize
\centering
\caption{Bias and root mean squared error (rMSE) for parameter estimates from 500 realizations of simulation Settings 1, 2, and 3. Estimates marked with a ``-'' are not obtained by the given estimation method.} \label{parameter-results-table}
\begin{threeparttable}
\begin{tabular}{clc rr rr rr rr rr rr}
\hline
\noalign{\vskip 0.2cm}
        & &       & \multicolumn{2}{c}{Direct Max.\tnote{1}}       &
        \multicolumn{2}{c}{EM\tnote{2}}& 
        \multicolumn{2}{c}{MCMC\tnote{2}}& 
        \multicolumn{2}{c}{\textit{SAMBA}-Specificity\tnote{3}}& 
        \multicolumn{2}{c}{\textit{SAMBA}-Sensitivity\tnote{4}}& 
        \multicolumn{2}{c}{Naive Analysis\tnote{5}}\\
        \cline{4-15}
Scenario &       & Truth & \multicolumn{1}{c}{Bias} & \multicolumn{1}{c}{rMSE} & \multicolumn{1}{c}{Bias} & \multicolumn{1}{c}{rMSE} & \multicolumn{1}{c}{Bias} & \multicolumn{1}{c}{rMSE} & \multicolumn{1}{c}{Bias} & \multicolumn{1}{c}{rMSE} & \multicolumn{1}{c}{Bias} & \multicolumn{1}{c}{rMSE} & \multicolumn{1}{c}{Bias} & \multicolumn{1}{c}{rMSE} \\
\hline
    \\
      & $\beta_0$ & 1 & 0.010 & 0.114 & 0.027 & 0.256 & -0.230 & 0.254 & 0.073 & 0.211 & -0.634 & 0.640 & -0.555 & 0.561 \\
(1)   & $\beta_X$ & -2 & -0.027 & 0.183 & -0.106 & 0.397 & 0.293 & 0.319 & 0.415 & 0.454 & 0.988 & 0.992 & 1.019 & 1.023 \\
      & $\gamma_{110}$ & 0.5 & -0.008 & 0.122 & -0.004 & 0.267 & 0.504 & 0.522 & 0.360 & 0.455 & - & - & - & -\\
      & $\gamma_{11Z}$ & 1 & 0.022 & 0.170 & 0.037 & 0.360 & -0.310 & 0.325 & -0.050 & 0.375 & - & - &  - & -\\
      & $\gamma_{120}$ & -0.5 & 0.011 & 0.244 & 0.095 & 0.810 & -0.651 & 0.660 & - & - & -2.754 & 2.900 & - & -\\
      & $\gamma_{12Z}$ & -1 & -0.055 & 0.300 & -0.319 & 1.464 & 0.518 & 0.523 & - & - & 0.894 & 0.994 & - & -\\
              \\
      & $\beta_0$ & 1 & 0.006 & 0.053 & 0.005 & 0.053 & -0.001 & 0.052 & 0.038 & 0.061 & -0.373 & 0.374 & -0.350 & 0.351\\
(2)   & $\beta_X$ & -2 & -0.010 & 0.089 & -0.012 & 0.091 & -0.041 & 0.077 & 0.183 & 0.191 & 0.628 & 0.629 & 0.643 & 0.643 \\
      & $\gamma_{110}$ & 0.5 & -0.003 & 0.155 & 0.007 & 0.153 & 0.223 & 0.253 & 0.239 & 0.284 & - & - & - & -\\
      & $\gamma_{11Z}$ & 1 & 0.004 & 0.120 & -0.004 & 0.118 & -0.138 & 0.159 & -0.031 & 0.125 & - & - & - & -\\
      & $\gamma_{120}$ & -0.5 & 0.020 & 0.442 & 0.015 & 0.429 & -0.532 & 0.555 & - & - & -3.656 & 3.699 & - & -\\
      & $\gamma_{12Z}$ & -1 & -0.045 & 0.314 & -0.036 & 0.306 & 0.306 & 0.316 & - & - & 0.871 & 0.888 & - & -\\
      \\
      & $\beta_0$ & 1 & 0.004 & 0.099 & -0.005 & 0.097 & -0.102 & 0.125 & -0.006 & 0.093 & -0.781 & 0.781 & -0.757 & 0.758\\
(3)   & $\beta_X$ & -2 & -0.031 & 0.188 & -0.040 & 0.123 & -0.083 & 0.119 & 0.006 & 0.099 & 0.831 & 0.832 & 0.844 & 0.845 \\
      & $\gamma_{110}$ & 0.5 & -0.010 & 0.111 & -0.024 & 0.109 & 0.055 & 0.109 & -0.009 & 0.105 & - & - & - & -\\
      & $\gamma_{11Z}$ & 1 & 0.010 & 0.135 & 0.018 & 0.133 & -0.068 & 0.106 & 0.019 & 0.131 & - & - & - & -\\
      & $\gamma_{120}$ & -5 &-2.320 & 5.900 & -3.094 & 28.586 & 3.042 & 3.043 & - & - & -1.260 & 1.389 & - & -\\
      & $\gamma_{12Z}$ & -5 & 2.480 & 8.190 & 4.023 & 9.406 & 4.302 & 4.304 & - & - & 5.787 & 5.791 & - & -\\
              \\
\hline  
\end{tabular}
\begin{tablenotes}
\item[1] ``Direct Max.'' estimates were obtained using the \texttt{optim} function in \texttt{R}. \cite{stats2021R}
\item[2] ``EM'' and ``MCMC'' estimates were computed using the \textit{COMBO} R Package. \cite{COMBO}
\item[3] The ``\textit{SAMBA}-Specificity'' results assume perfect specificity and were computed using the \textit{SAMBA} R Package.
\item[4] The ``\textit{SAMBA}-Sensitivity'' results were computed using an adapted function from the \textit{SAMBA} R Package.  \citep{beesley2020statistical}
\item[5]  The ``Naive Analysis'' results were obtained by running a simple logistic regression model for $Y^* \sim \boldsymbol{X}$.
\end{tablenotes}
\end{threeparttable}
\end{table}

\clearpage

\begin{table}[H]
\centering
\caption{Estimated event probabilities from 500 realizations of simulation Settings 1, 2, and 3. Cells with a bold 1 (denoted \textbf{1}) indicate that a given procedure fixed that value at 1. } \label{probability-results-table}
\begin{threeparttable}
\begin{tabular}{ccrccccccc}
\hline
\noalign{\vskip 0.2cm}
        Scenario & & & Data\tnote{1} & Direct Max.\tnote{2} & EM\tnote{3}& MCMC \tnote{3}& SAMBA-Specificity \tnote{4}& SAMBA-Sensitivity \tnote{5}& Naive Analysis \tnote{6}\\
\hline
    \\
(1)           && $P(Y = 1)$ & 0.647 & 0.648 & 0.646 & 0.625 & 0.676 & 0.574 & 0.590 \\
              && $P(Y = 2)$ & 0.353 & 0.352 & 0.354 & 0.375 & 0.324 & 0.426 & 0.410 \\
              && $P(Y^* = 1 | Y = 1)$ & 0.847 & 0.858 & 0.856 & 0.874 & 0.883 & \textbf{1} & \textbf{1} \\
              && $P(Y^* = 2 | Y = 2)$ & 0.877 & 0.859 & 0.853 & 0.862 & \textbf{1} & 0.961 & \textbf{1} \\
              \\
              \\
(2)           && $P(Y = 1)$ & 0.648 & 0.648 & 0.648 & 0.646 & 0.661 & 0.614 & 0.618 \\
              && $P(Y = 2)$ & 0.352 & 0.352 & 0.352 & 0.354 & 0.339 & 0.386 & 0.382 \\
              && $P(Y^* = 1 | Y = 1)$ & 0.924 & 0.931 & 0.930 & 0.930 & 0.941 & \textbf{1} & \textbf{1} \\
              && $P(Y^* = 2 | Y = 2)$ & 0.945 & 0.931 & 0.928 & 0.928 & \textbf{1} & 0.988 & \textbf{1} \\
              \\
              \\
(3)           && $P(Y = 1)$ & 0.647 & 0.647 & 0.645 & 0.630 & 0.647 & 0.543 & 0.548 \\
              && $P(Y = 2)$ & 0.353 & 0.353 & 0.355 & 0.370 & 0.353 & 0.457 & 0.452 \\
              && $P(Y^* = 1 | Y = 1)$ & 0.846 & 0.857 & 0.856 & 0.855 & 0.858 & \textbf{1} & \textbf{1} \\
              && $P(Y^* = 2 | Y = 2)$ & 1.000 & 0.995 & 0.991 & 0.946 & \textbf{1} & 0.990 & \textbf{1} \\
              \\
\hline  
\end{tabular}
\begin{tablenotes}
\item[1] ``Data'' terms refer to empirical values computed from generated datasets.
\item[2] ``Direct Max.'' estimates were obtained using the \texttt{optim} function in \texttt{R}. \cite{stats2021R}
\item[3] ``EM'' and ``MCMC'' estimates were computed using the \textit{COMBO} R Package. \cite{COMBO}
\item[4] The ``\textit{SAMBA}-Specificity'' results assume perfect specificity and were computed using the \textit{SAMBA} R Package.
\item[5] The ``\textit{SAMBA}-Sensitivity'' results were computed using an adapted function from the \textit{SAMBA} R Package.  \citep{beesley2020statistical}
\item[6]  The ``Naive Analysis'' results were obtained by running a simple logistic regression model for $Y^* \sim \boldsymbol{X}$.
\end{tablenotes}
\end{threeparttable}
\end{table}

\newpage

\begin{table}[H]
\centering
\caption{Parameter estimates and standard errors from the applied example using the MEPS dataset. All methods are described in Table \ref{parameter-results-table}. Estimates marked with a ``-'' are not obtained by the given estimation method.} \label{applied-results-table}
\begin{threeparttable}
\begin{tabular}{ll rr rr rr rr rr rr}
\hline
           && \multicolumn{2}{c}{Direct Max.}       &
           \multicolumn{2}{c}{EM}       &
        \multicolumn{2}{c}{MCMC}       &
         \multicolumn{2}{c}{\textit{SAMBA}-Specificity} & 
        \multicolumn{2}{c}{\textit{SAMBA}-Sensitivity} & 
        \multicolumn{2}{c}{Naive Analysis}\\
        \cline{3-14}
 && \multicolumn{1}{c}{Est.} & \multicolumn{1}{c}{SE} & \multicolumn{1}{c}{Est.} & \multicolumn{1}{c}{SE} & \multicolumn{1}{c}{Est.} & \multicolumn{1}{c}{SE} & \multicolumn{1}{c}{Est.} & \multicolumn{1}{c}{SE} & \multicolumn{1}{c}{Est.} & \multicolumn{1}{c}{SE} & \multicolumn{1}{c}{Est.} & \multicolumn{1}{c}{SE}\\
\hline
    \\
      $\beta_0$\tnote{4} && -4.376 & 0.603 & -4.374 & 0.065 & -3.937 & 0.213 & -3.184 & 0.665 & -4.159 & 101.296 & -3.576 & 0.078 \\
    $\beta_{smoke}$\tnote{1} && 1.545 & 0.370 & 1.544 & 0.107 & 1.070 & 0.222 & 0.643 & 0.128 & 0.763 & 3.256 & 0.635 & 0.109 \\
    $\beta_{exercise}$\tnote{2} && 0.304 & 0.273 & 0.303 & 0.126 & 0.279 & 0.262 & 0.273 & 0.095 & 0.428 & 1.476 & 0.184 & 0.084 \\
    $\beta_{age}$\tnote{3} && 0.094 & 0.014 & 0.094 & 0.010 & 0.019 & 0.074 & 0.065 & 0.009 & 0.058 & 2.688 & 0.059 & 0.003 \\
       $\gamma_{110}$\tnote{4} && 2.975 & 2.048 & 2.969 & 0.100 & 0.973 & 0.311 & 2.437 & 7.913 & - & - & - & -\\
       $\gamma_{11,gender}$\tnote{5} && -1.768 & 1.003 & -1.766 & 0.036 & -0.142 & 0.315 & -2.661 & 6.750 & - & - &  - & -\\
       $\gamma_{11,age}$\tnote{6} && -0.199  & 0.092 & -0.198 & 0.005 & 0.107 & 0.243 & -0.005 & 0.012 & - & - &  - & -\\
       $\gamma_{120}$\tnote{4} && -3.579 & 0.264 & -3.580 & 0.112 & -3.296 & 0.122 & - & - & -3.674 & 102.790 & - & -\\
       $\gamma_{12,gender}$\tnote{5} && -0.818 & 0.151 & -0.818 & 0.108 & -0.975 & 0.124 & - & - & -5.542 & 5.206 & - & -\\
       $\gamma_{12,age}$\tnote{6} && 0.083 & 0.012 & 0.084 & 0.005 & 0.072 & 0.010 & - & - & 0.063 & 2.678 & - & -\\
              \\
\hline  
\end{tabular}
\begin{tablenotes}
\item[1] $\beta_{smoke}$ refers to the association between smoking status (reference = non-smoker) and myocardial infarction (MI).
\item[2] $\beta_{exercise}$ refers to the association between exercise habits (reference = regular exercise) and MI.
\item[3] $\beta_{age}$ refers to the association between age (centered and scaled) and MI.
\item[4] $\beta_0$, $\gamma_{110}$, and $\gamma_{120}$ are intercept terms.
\item[5] $\gamma_{1,j,gender}$ refers to the association between gender and observed MI, given true MI status $j$ (reference = man).
\item[6] $\gamma_{1,j,age}$ refers to the association between age and observed MI, given true MI status $j$.
\end{tablenotes}
\end{threeparttable}
\end{table}

\end{landscape}

\newpage

\begin{appendices}

\section{Starting Values for Direct Optimization of the Observed Data Log-likelihood} \label{optim-starting-values}
This section suggests strategies for parameter selecting starting values for direct optimization of the observed data log-likelihood (using the \texttt{optim()} or \texttt{nlm()} function in \texttt{R}). In general, starting values for $\boldsymbol{\beta}$ parameters should be obtained from the naive \textit{analysis model}, $Y^* | \boldsymbol{X}$. We recommend starting all $\boldsymbol{\gamma}$ parameters at values that produce reasonable sensitivity and specificity estimates for the analysis dataset and context, but are small in magnitude.

For settings with moderate sample size (i.e. in the tens of thousands) and with a small number of covariates (i.e. a single predictor in both the true outcome and observation mechanisms), direct optimization is a feasible and stable option. For datasets with even a moderate number of covariates (i.e. an $n$ x $3$ $\boldsymbol{X}$ matrix and an $n$ x $2$ $\boldsymbol{Z}$ matrix), we found built-in optimization functions to be sensitive to starting values, particularly for $\boldsymbol{\gamma}$ terms, leading to unstable results. In these cases, analysts may consider different starting values for $\boldsymbol{\gamma_{kj0}}$ terms (intercepts) and $\boldsymbol{\gamma_{kjZ}}$ terms (slopes). For example, a reasonable set of starting values to yield a ``null'' starting set for $\boldsymbol{\gamma}$ parameters might be: a moderate positive value for $\gamma_{110}$ (i.e. 1 to 3), a moderate negative value for $\gamma_{120}$ (i.e. -3 to -1), and 0 for all $\boldsymbol{\gamma_{kjZ}}$ terms. This starting value setting corresponds to a case with low misclassification (i.e. high sensitivity and high specificity) and $\boldsymbol{Z}$ covariates that have no association with the probability of misclassification. Analysts should also consider $\boldsymbol{\gamma}$ starting values that yield the expected correspondence between sensitivity and specificity. For example, if it is expected that sensitivity will be higher than specificity in a given context, choosing starting values such that $\pi^*_{11} > \pi^*_{22}$ would be prudent. It is important to note that \textit{a priori} adjustment of $\boldsymbol{\gamma}$ starting values, as described, does not make direct maximization of the observed data log-likelihood a more stable procedure, generally. Rather, these starting values tend to ensure convergence closer to the results obtained from other, more stable, estimation methods like those described in Sections \ref{em} and \ref{mcmc}, respectively.

\section{Details on Label Switching Corrections}
\subsection{Sensitivity and Specificity Estimates under Label Switching}\label{appendix-label-switch}
This section details the relationship between the estimated sensitivity and specificity from $\boldsymbol{\hat{\beta}}$ and $\boldsymbol{\hat{\gamma}}$, denoted $\hat{\pi}^*_{11}$ and $\hat{\pi}^*_{22}$ respectively, and the estimated sensitivity and specificity from the label-switched parameter set, denoted $\hat{\pi}^{*, switch}_{11}$ and $\hat{\pi}^{*, switch}_{22}$ respectively. The relationship is provided in (\ref{eq:misclassification_label_switch}) as follows: $\hat{\pi}^{*, switch}_{11} = 1 - \hat{\pi}^*_{22}$ and $\hat{\pi}^{*, switch}_{22} = 1 - \hat{\pi}^*_{11}$.

To show this relationship, first define the estimated average correct classification rates using parameter estimates $\boldsymbol{\hat{\beta}}$ and $\boldsymbol{\hat{\gamma}}$, obtained from a method in Section \ref{estimation-methods}. These estimated averages are taken across all observations, for both $j = 1$ and $j = 2$, as shown in (\ref{eq:avg_pistar}) and rewritten here: $\hat{\pi}^*_{jj} = \sum_{i = 1}^N \hat{\pi}^*_{ijj} = \sum_{i = 1}^N \frac{\text{exp}\{\hat{\gamma}_{jj0} + \boldsymbol{\hat{\gamma}_{jjZ} Z_i}\}}{1 + \text{exp}\{\hat{\gamma}_{jj0} + \boldsymbol{\hat{\gamma}_{jjZ} Z_i\}}}$.

Denote the label-switched estimated parameter set as $\boldsymbol{\hat{\beta}}^{switch}$ and $\boldsymbol{\hat{\gamma}}^{switch}$. Note that $\boldsymbol{\hat{\beta}}^{switch} = -\boldsymbol{\hat{\beta}}$, $\boldsymbol{\hat{\gamma}_{k1}}^{switch} = \boldsymbol{\hat{\gamma}_{k2}}$, and $\boldsymbol{\hat{\gamma}_{k2}}^{switch} = \boldsymbol{\hat{\gamma}_{k1}}$, as described in Section \ref{label-switching}. Average correct classification rates, denoted $\hat{\pi}^{*, switch}_{jj}$, can also be estimated using $\boldsymbol{\hat{\beta}}^{switch}$ and $\boldsymbol{\hat{\gamma}}^{switch}$. These averages are also taken across all observations, for both $j = 1$ and $j = 2$, as follows: 
\begin{equation}
\begin{aligned}
\label{eq:avg_pistar_switch}
\hat{\pi}^{*, switch}_{jj} = \sum_{i = 1}^N \hat{\pi}^{*, switch}_{ijj} = \sum_{i = 1}^N \frac{\text{exp}\{\hat{\gamma}^{switch}_{jj0} + \boldsymbol{\hat{\gamma}^{switch}_{jjZ} Z_i} \}}{1 + \text{exp}\{\hat{\gamma}^{switch}_{jj0} + \boldsymbol{\hat{\gamma}^{switch}_{jjZ} Z_i} \}}.
\end{aligned}
\end{equation}

First, suppose $j = 1$. For $j = 1$, we have
\begin{equation}
\begin{aligned}
\label{eq:avg_pistar_switch11}
\hat{\pi}^{*, switch}_{11} = \sum_{i = 1}^N \hat{\pi}^{*, switch}_{i11} = \sum_{i = 1}^N \frac{\text{exp}\{\hat{\gamma}^{switch}_{110} + \boldsymbol{\hat{\gamma}^{switch}_{11Z} Z_i}\}}{1 + \text{exp}\{\hat{\gamma}^{switch}_{110} + \boldsymbol{\hat{\gamma}^{switch}_{11Z} Z_i}\}}.
\end{aligned}
\end{equation}

Since $\boldsymbol{\hat{\gamma}_{11}}^{switch} = \boldsymbol{\hat{\gamma}_{12}}$, we can write: 
\begin{flalign}
\begin{aligned}
\label{eq:avg_pistar_switch11}
\hat{\pi}^{*, switch}_{11} &= \sum_{i = 1}^N \hat{\pi}^{*, switch}_{i11} = \sum_{i = 1}^N \frac{\text{exp}\{\hat{\gamma}^{switch}_{110} + \boldsymbol{\hat{\gamma}^{switch}_{11Z} Z_i}\}}{1 + \text{exp}\{\hat{\gamma}^{switch}_{110} + \boldsymbol{\hat{\gamma}^{switch}_{11Z} Z_i} \}} \\
& = \sum_{i = 1}^N \frac{\text{exp}\{\hat{\gamma}_{120} + \boldsymbol{\hat{\gamma}_{12Z} Z_i}\}}{1 + \text{exp}\{\hat{\gamma}_{120} + \boldsymbol{\hat{\gamma}_{12Z} Z_i} \}} \\
& = \sum_{i = 1}^N \Bigl[ 1 - \frac{1}{1 + \text{exp}\{\hat{\gamma}_{120} + \boldsymbol{\hat{\gamma}_{12Z} Z_i}\}} \Bigr] \\
& = \sum_{i = 1}^N \Bigl[ 1 - \hat{\pi}^*_{i22}\Bigr] \\
&= 1 - \hat{\pi}^*_{22}.
\end{aligned}
\end{flalign}

Next, suppose $j = 2$. For $j = 2$, we have 
\begin{equation}
\begin{aligned}
\label{eq:avg_pistar_switch11}
\hat{\pi}^{*, switch}_{22} = \sum_{i = 1}^N \hat{\pi}^{*, switch}_{i22} = \sum_{i = 1}^N \frac{1}{1 + \text{exp}\{\hat{\gamma}^{switch}_{120} + \boldsymbol{\hat{\gamma}^{switch}_{12Z} Z_i}\}}.
\end{aligned}
\end{equation}

Since $\boldsymbol{\hat{\gamma}_{12}}^{switch} = \boldsymbol{\hat{\gamma}_{11}}$, we can write: 
\begin{flalign}
\begin{aligned}
\label{eq:avg_pistar_switch22}
\hat{\pi}^{*, switch}_{22} &= \sum_{i = 1}^N \hat{\pi}^{*, switch}_{i22} = \sum_{i = 1}^N \frac{1}{1 + \text{exp}\{\hat{\gamma}^{switch}_{120} + \boldsymbol{\hat{\gamma}^{switch}_{12Z} Z_i} \}} \\
& = \sum_{i = 1}^N \frac{1}{1 + \text{exp}\{\hat{\gamma}_{110} + \boldsymbol{\hat{\gamma}_{11Z} Z_i}\}} \\
& = \sum_{i = 1}^N \Bigl[ 1 - \frac{\text{exp}\{\hat{\gamma}_{110} + \boldsymbol{\hat{\gamma}_{11Z} Z_i} \}}{1 + \text{exp}\{\hat{\gamma}_{110} + \boldsymbol{\hat{\gamma}_{11Z} Z_i}\}} \Bigr] \\
& = \sum_{i = 1}^N \Bigl[ 1 - \hat{\pi}^*_{i11}\Bigr] \\
&= 1 - \hat{\pi}^*_{11}.
\end{aligned}
\end{flalign}
Thus we have shown the relationship in (\ref{eq:misclassification_label_switch}).

Note that our criteria to correct for label switching utilizes a comparison between values of Youden's $J$ Statistic for each parameter set. We suggest selecting the parameter set with the larger value of Youden's $J$ Statistic. Equivalently, we are selecting the parameter set that jointly maximizes sensitivity and specificity. Next, we briefly demonstrate that this criteria is equivalent to requiring that the sum of sensitivity and specificity exceed $1$. 

Suppose that the sum of the sensitivity and specificity for a given parameter set, $\pi^*_{11}$ + $\pi^*_{22}$, is greater than $1$. This would suggest that the sum of the sensitivity and specificity of the label switched parameter set is \textit{not} greater than $1$:
\begin{flalign}
\begin{aligned}
\label{eq:avg_pistar_switch22}
\hat{\pi}^*_{11} + \hat{\pi}^*_{22} > 1 &\implies (1 -  \hat{\pi}^{*, switch}_{11}) + (1 - \hat{\pi}^{*, switch}_{22}) > 1 \\
&\implies 2 - (\hat{\pi}^{*, switch}_{11} + \hat{\pi}^{*, switch}_{22}) > 1 \\
&\implies 1 > \hat{\pi}^{*, switch}_{11} + \hat{\pi}^{*, switch}_{22}
\end{aligned}
\end{flalign}
Thus, selecting the parameter set such that has the larger Youden's $J$ Statistic, defined as $\hat{\pi}^*_{11} + \hat{\pi}^*_{22} - 1$, is equivalent to selecting the parameter set such that the sum of the sensitivity and specificity is greater than $1$. 

\subsection{Algorithm for Label Switching Correction}\label{appendix-algorithm}

This section provides an algorithm and pseudocode to aid in the explanation of the procedure to correct label switching described in Section \ref{label-switching-correction}. 

\begin{algorithm}[H]
\caption{Correcting label switching in binary outcome misclassification models}\label{alg:label-switch}
\begin{algorithmic}
\State Compute $\hat{J}$ and $\hat{J}^{switch}$ using $\boldsymbol{\hat{\beta}}$ and $\boldsymbol{\hat{\gamma}}$.
\If{$\hat{J} \geq \hat{J}^{switch}$}
    \State $\boldsymbol{\hat{\beta}}_{corrected} \gets \boldsymbol{\hat{\beta}}$
    \State $\boldsymbol{\hat{\gamma}}_{corrected} \gets \boldsymbol{\hat{\gamma}}$
\Else
    \State $\boldsymbol{\hat{\beta}}_{corrected} \gets -\boldsymbol{\hat{\beta}}$
    \State $\boldsymbol{\hat{\gamma}}_{corrected, k1} \gets \boldsymbol{\hat{\gamma}}_{k2}$
    \State $\boldsymbol{\hat{\gamma}}_{corrected, k2} \gets \boldsymbol{\hat{\gamma}}_{k1}$
\EndIf 
\end{algorithmic}
\end{algorithm}

\subsection{Label Switching Correction of the Information Matrix} \label{appendix-info-matrix-switch}
We present an example information matrix to illustrate the label switching correction described in Section \ref{label-switching-correction}. Consider an example where, for each $y_{ij}$, we have a single predictor $x_{i}$ in the true outcome mechanism and a single predictor $z_i$ in the observation mechanisms, as in Section \ref{permutation-invariance-complete}. In such a case, label switching produces two parameter sets: (1) $(\beta_0, \beta_X, \gamma_{110},  \gamma_{11Z},$ 
$\gamma_{120}, \gamma_{12Z})$,
and
(2) $(-\beta_0, -\beta_X, \gamma_{120}, \gamma_{12Z}, \gamma_{110}, \gamma_{11Z})$. If $\hat{J} < \hat{J}^{switch}$, the ``label-switched'' parameter set is deemed to be the appropriate parameter set, and the information matrix must be adjusted accordingly. The following equation demonstrates the correspondence between the original information matrix and original parameters and the information matrix and parameters that are corrected for label switching. The parameters listed above the matrix rows and columns label which entries of the matrix correspond to which parameter name. The $d$ and $r$ values denote arbitrary diagonal and non-diagonal real-valued entries in the information matrix, respectively.

\begin{equation}
\begin{blockarray}{ccccccc}
\beta_0 & \beta_X & \gamma_{110} & \gamma_{11Z} & \gamma_{120} & \gamma_{12Z}\\
\begin{block}{(cccccc)c}
  d_1 & r_1 & r_2 & r_3 & r_4 & r_5 & \beta_0 \\
  r_1 & d_2 & r_6 & r_7 & r_8 & r_9 & \beta_X \\
  r_2 & r_6 & d_3 & r_{10} & r_{11} & r_{12} & \gamma_{110} \\
  r_3 & r_7 & r_{10} & d_4 & r_{13} & r_{14} & \gamma_{11Z} \\
  r_4 & r_8 & r_{11} & r_{13} & d_5 & r_{15} & \gamma_{120} \\
  r_5 & r_9 & r_{12} & r_{14} & r_{15} & d_6 & \gamma_{12Z} \\
\end{block}
\\
     &&&    \longrightarrow
\vspace{1em}
\\
-\beta_0 & -\beta_X & \gamma_{120} & \gamma_{12Z} & \gamma_{110} & \gamma_{11Z}\\
\begin{block}{(cccccc)c}
  d_1 & r_1 & r_2 & r_3 & r_4 & r_5 & -\beta_0 \\
  r_1 & d_2 & r_6 & r_7 & r_8 & r_9 & -\beta_X \\
  r_2 & r_6 & d_3 & r_{10} & r_{11} & r_{12} & \gamma_{120} \\
  r_3 & r_7 & r_{10} & d_4 & r_{13} & r_{14} & \gamma_{12Z} \\
  r_4 & r_8 & r_{11} & r_{13} & d_5 & r_{15} & \gamma_{110} \\
  r_5 & r_9 & r_{12} & r_{14} & r_{15} & d_6 & \gamma_{11Z} \\
\end{block}
\end{blockarray}
\end{equation}

\section{Simulation Study Settings} \label{appendix-sim}

\subsection{Simulation Settings} \label{appendix-sim-settings}
We present simulations for evaluating the proposed binary outcome misclassification model in terms of bias and root mean squared error (rMSE). For a given simulation scenario, we present parameter estimates for a binary outcome misclassification model obtained from the EM algorithm and from MCMC. We compare these estimates to the naive \textit{analysis model} that considers only potentially misclassified outcomes $Y^*$ and predictors $\boldsymbol{X}$. In addition, we use the \textit{SAMBA} R package from Beesley and Mukherjee to find the relevant parameter estimates in the event that we assumed that our \textit{observation mechanism} had perfect specificity.\cite{beesley2020statistical} We also consider the case where we incorrectly assumed that our \textit{observation mechanism} had perfect sensitivity.

For the EM analysis, we used the "squarem" acceleration scheme from the \texttt{turboEM} package in \texttt{R}. \cite{turboEM, stats2021R} The convergence criterion for the algorithm required that the absolute difference between successive log-likelihood values was less than the tolerance of $1 \times 10^{-7}$. The alternative stopping criteria was reached when the number of iterations surpassed 1500. For the perfect specificity scenario in Setting 3, we found that the number of iterations of the EM algorithm required for convergence was generally higher than in other settings (i.e. $\approx 5,000$).

For the MCMC analysis, prior specifications are provided in Table \ref{sim-setting-table1}. All priors for $\boldsymbol{\beta}$ coefficients were Normal distributions, centered at estimates from the naive \textit{analysis model} with a standard deviation of 10. For the intercept of the sensitivity portion of the observation mechanism, $\gamma_{110}$, we specified a \textit{Normal}$(1, 10)$ prior. For the intercept of the specificity mechanism, $\gamma_{120}$, we specified a \textit{Normal}$(1, 10)$ prior. All remaining $\boldsymbol{\gamma}$ slope parameters had a \textit{Normal}$(0, 10)$ prior. The means of the prior distributions for the $\boldsymbol{\gamma}$ terms correspond to prior sensitivity and specificity estimates of 0.731 each. In addition, the 0 prior means of the  $\boldsymbol{\gamma}$ slope parameters correspond to a case where the $z$ covariate is unrelated to the misclassification mechanism. The selection of these prior means follows identical logic to the choice of starting values for $\boldsymbol{\beta}$ and $\boldsymbol{\gamma}$ coefficients for direct optimization of the observed data log-likelihood (Appendix \ref{optim-starting-values}). A relatively large standard deviation was selected for each prior distribution in order to reflect some uncertainty in the choice of prior means. For all simulation settings, we ran $4$ Markov chains of length $5,000$, with a burn-in of length $2,000$. 

In all settings, we generate $500$ datasets with $P(Y = 1) \approx 65\%$. In the first simulation setting, we studied an example that is expected to be highly problematic for analysts: a relatively small sample size and relatively high misclassification rate. In this setting, we generated datasets with $1000$ members and imposed outcome misclassification rates between $10\%$ and $18\%$. In the Setting 2, we show that the problem of outcome misclassification remains influential even as sample size increases and misclassification rates decrease. In this setting, we generated datasets with $10000$ members and imposed outcome misclassifiation rates between $5\%$ and $10\%$. In the third simulation setting, we consider a case with perfect specificity. The purpose of this scenario is to demonstrate that our methods recover unbiased association parameters, even in cases where we unnecessarily account for bidirectional misclassification. In this setting, we generated datasets with $5000$ members and imposed sensitivity rates around $82\%$ to $90\%$.

For a dataset with $1000$ members, the analysis using our proposed EM algorithm took about $17$ seconds while our proposed MCMC analysis took approximately $35$ minutes. It is possible for analysts to parallelize MCMC chains to improve computation time. Direct optimization using the \texttt{optim()} function in \texttt{R} took approximately 6 seconds. \citep{stats2021R}

These settings are outlined in Table \ref{sim-setting-table1}. All analyses were conducted in \texttt{R}. \citep{stats2021R}

\begin{table}[htbt]
\centering
\caption{Number of generated datasets (N. Realizations), Sample size ($N$), $P(Y = 1)$, $P(Y^* = 1 | Y = 1)$ (sensitivity), $P(Y^* = 2 | Y = 2)$ (specificity), $\boldsymbol{\beta}$ prior distribution, and $\boldsymbol{\gamma}$ prior distribution settings for each of the the simulation Settings 1, 2, and 3.} \label{sim-setting-table1}
\begin{threeparttable}
\begin{tabular}{clrrrrr}
\hline
        Scenario & & & & Setting \\
\hline
    \\
(1)           && N. Realizations & & 500\\
              && $N$ & & 1000\\
              && $P(Y = 1)$ & & 0.65 \\
              && $P(Y^* = 1 | Y = 1)$ && 0.82 - 0.90 \\
              && $P(Y^* = 2 | Y = 2)$ && 0.82 - 0.90 \\
              && $\boldsymbol{\beta}$ prior distributions\tnote{1} && \textit{Normal}$(\hat{\boldsymbol{\beta}}^{naive}, 10)$ \\
              && $\gamma_{110}$ prior distribution && \textit{Normal}$(1, 10)$ \\
              && $\gamma_{120}$ prior distribution && \textit{Normal}$(-1, 10)$ \\
              && $\boldsymbol{\gamma}_{kjZ}$ prior distributions && \textit{Normal}$(0, 10)$ \\
              \\
              \\
(2)           && N. Realizations & & 500\\
              && $N$ & & 10000\\
              && $P(Y = 1)$ & & 0.65 \\
              && $P(Y^* = 1 | Y = 1)$ && 0.90 - 0.95 \\
              && $P(Y^* = 2 | Y = 2)$ && 0.90 - 0.95 \\
              && $\boldsymbol{\beta}$ prior distributions\tnote{1} && \textit{Normal}$(\hat{\boldsymbol{\beta}}^{naive}, 10)$ \\
              && $\gamma_{110}$ prior distribution && \textit{Normal}$(1, 10)$ \\
              && $\gamma_{120}$ prior distribution && \textit{Normal}$(-1, 10)$ \\
              && $\boldsymbol{\gamma}_{kjZ}$ prior distributions && \textit{Normal}$(0, 10)$ \\
              \\
              \\
(3)           && N. Realizations & & 500\\
              && $N$ & & 5000\\
              && $P(Y = 1)$ & & 0.65 \\
              && $P(Y^* = 1 | Y = 1)$ && 0.82 - 0.90 \\
              && $P(Y^* = 2 | Y = 2)$ && 1 \\
              && $\boldsymbol{\beta}$ prior distributions\tnote{1} && \textit{Normal}$(\hat{\boldsymbol{\beta}}^{naive}, 10)$ \\
              && $\gamma_{110}$ prior distribution && \textit{Normal}$(1, 10)$ \\
              && $\gamma_{120}$ prior distribution && \textit{Normal}$(-1, 10)$ \\
              && $\boldsymbol{\gamma}_{kjZ}$ prior distributions && \textit{Normal}$(0, 10)$ \\
              \\
\hline  
\end{tabular}
\begin{tablenotes}
\item[1] $\hat{\boldsymbol{\beta}}^{naive}$ refers to the vector of $\hat{\boldsymbol{\beta}}$ coefficient estimates obtained from the ``Naive Analysis'' method, where a simple logistic regression model is fit for $Y^* \sim \boldsymbol{X}$
\end{tablenotes}
\end{threeparttable}
\end{table}

\subsection{Data Generation}\label{appendix-sim-data-gen}
For each of the simulated datasets, we begin by generating the predictors $X$ and $Z$ from a multivariate Normal distribution. In Settings 1 and 3, the means were 0 and 1.5, respectively, for $X$ and $Z$. In Setting 2, the means were 0 and 2.5, respectively, for $X$ and $Z$. Covariate generation in all simulation settings used unit variances and covariance terms equal to 0.30. The absolute value of all generated $\boldsymbol{z}$ terms was taken.

Next, we generated true outcome status using the following relationship: $P(Y = 1 | X) = 1 + (-2)X$. For Settings 1 and 2, we used the following relationships to obtain $Y^*$: $P(Y^* = 1 | Y = 1, Z) = 0.50 + (1)Z$ and $P(Y^* = 1 | Y = 2, Z) = -0.50 + (-1)Z$. The different $Z$ distribution between Settings 1 and 2 resulted in different misclassification rates. In Setting 1, misclassification rates were between $10\%$ and $18\%$. In Setting 2, misclassification rates were between $5\%$ and $10\%$. In Setting 3, we generated $Y^*$ using the following relationships: $P(Y^* = 1 | Y = 1, Z) = 0.50 + (1)Z$ and $P(Y^* = 1 | Y = 2, Z) = -5 + (-5)Z$. The choice of parameter values $(-5, -5)$ for $P(Y^* = 1 | Y = 2)$ resulted in near perfect specificity in the generated datasets. In the simulation study presented in Section \ref{sim-setting-table1}, the average specificity in the generated data for Setting 3 was 1.000 (see Table \ref{probability-results-table}). The minimum specificity in the Setting 3 generated samples was 0.997 and the maximum specificity in the Setting 3 generated samples was 1.000. Only 30 of the 500 generated datasets had true specificity less than 0.999. 

\subsection{Simulation Results for Alternate Prior Selection}
As discussed in Collins and Huynh (2014), \cite{collins2014estimation} the elucidation of prior distributions can be challenging for misclassification models. Our main simulation results from the proposed MCMC method are obtained using Normal prior distributions for both the $\boldsymbol{\beta}$ and $\boldsymbol{\gamma}$ parameters. The specification of these Normal priors is described in detail in Section \ref{appendix-sim-settings}. Here, we present simulation results under a different prior distribution setting. In this case, we use $\textit{Uniform}(-10, 10)$ priors for all $\boldsymbol{\beta}$ and $\boldsymbol{\gamma}$ parameters. While the Uniform distribution is flat, and therefore noninformative across its range, using Uniform priors results in a restricted parameter space between the distribution's upper and lower bounds. 

Bias and rMSE for parameter estimates for all simulation scenarios are presented in Table \ref{appendix-mcmc-table} for the proposed MCMC analysis with both Uniform prior distributions and with Normal prior distributions for all parameters. Table \ref{appendix-mcmc-prob-table} displays estimated response and misclassification probabilities for each prior distribution and simulation setting. The results from the MCMC analysis with the Normal prior distributions are identical to those presented in the main text. They are recopied in these tables for ease of comparison.

In Setting 1, we see that bias is typically lower for each parameter when Uniform priors are used, compared to Normal priors (Table \ref{appendix-mcmc-table}). The lower bias in the Uniform prior analysis also resulted in more accurate estimation of the response probabilities, $P(Y = 1)$ and $P(Y = 2)$, compared to the Normal prior analysis. However, the Uniform prior estimates for $\gamma_{120}$ and $\gamma_{12Z}$ have larger bias than that from the Normal distribution prior setting. This finding is likely an artifact of the simulation setting. The fact that $P(Y = 2)$ is considerably smaller than $P(Y = 1)$ means that there were fewer observations were subject to imperfect specificity. Thus, the parameters governing the specificity component of the observation mechanism, $\gamma_{120}$ and $\gamma_{12Z}$, are generally been more difficult to estimate without an informative prior. In addition, the rMSE for all parameter estimates is typically lower for MCMC with Normal priors than for MCMC with Uniform priors. The relative flatness of the Uniform priors likely allowed for a greater variance in the parameter estimates, within the defined range of the distribution, compared to those obtained using Normal priors. The central peak of the Normal priors, however, likely contributed to the bias observed in the corresponding analysis, because parameter estimates were drawn toward an incorrect prior mean. 

With a larger sample size in Setting 2, we see that the characteristics of the prior distributions have less of an impact on the final estimates. Bias and rMSE are largely comparable between the analyses using Uniform priors and using Normal priors (Table \ref{appendix-mcmc-table}). Once again, the Uniform prior setting typically results in estimates with lower bias and higher rMSE than that from the Normal prior setting. This pattern, however, is less apparent than in Setting 1. Notably, the estimation of $\gamma_{120}$ and $\gamma_{12Z}$ is much improved for the Uniform prior analysis in Setting 2 compared to Setting 1, indicating that the larger sample size likely allowed for more accurate estimation of the parameters, despite the fact that $P(Y = 2)$ is still considerably smaller than $P(Y = 1)$ (Table \ref{appendix-mcmc-prob-table}). 

In Setting 3, we again see comparable bias and rMSE between the prior distribution settings, with the exception of $\gamma_{120}$ and $\gamma_{12Z}$ (Table \ref{appendix-mcmc-table}). The bias and rMSE for the estimates of $\gamma_{120}$ and $\gamma_{12Z}$ under the Uniform prior setting are much better than that under the Normal prior setting. It appears that the flatness of the Uniform prior allowed for more extreme parameter estimates compared to the Normal prior, which may have pulled estimates too strongly toward the prior mean. This pattern had a notable impact on specificity estimates. In Table \ref{appendix-mcmc-prob-table}, we see that the Uniform prior distribution analysis resulted in an unbiased estimate of $P(Y^* = 2 | Y = 2)$, while the Normal prior analysis would have led an analyst to assume that the specificity in Setting 3 may have been imperfect. 

The comparison of these prior distributions suggests that, Uniform priors may result in estimates with smaller bias and larger variance than those obtained with Normal priors. In cases of perfect sensitivity or specificity, or generally cases with extreme parameter values, the flat Uniform prior may allow for more accurate estimation than a Normal prior, even with a large standard deviation. However, analysts must be careful to ensure that the range of the Uniform prior does not restrict the parameter space such that reasonable parameter values are excluded.

\newpage

\begin{table}[H]\footnotesize
\centering
\caption{Bias and root mean squared error (rMSE) for parameter estimates from 500 realizations of simulation Settings 1, 2, and 3. Estimates were computed using the \textit{COMBO} R Package. \cite{COMBO}}\label{appendix-mcmc-table}
\begin{threeparttable}
\begin{tabular}{clc rr rr}
\hline
\noalign{\vskip 0.2cm}
        & &    &
        \multicolumn{2}{c}{MCMC} &
        \multicolumn{2}{c}{MCMC}\\
        &&& 
        \multicolumn{2}{c}{Uniform priors\tnote{1}} &
        \multicolumn{2}{c}{Normal priors\tnote{2}}\\
        \cmidrule{4-7}
Scenario &  & Truth & Bias & rMSE & Bias & rMSE\\
\hline
    \\
      & $\beta_0$ & 1 & 0.006 & 0.300 & -0.230 & 0.254\\
(1)   & $\beta_X$ & -2 & 0.154 & 0.530 & 0.293 & 0.319\\
      & $\gamma_{110}$ & 0.5 & 0.051 & 0.355 & 0.504 & 0.522\\
      & $\gamma_{11Z}$ & 1 & 0.297 & 0.844 & -0.310 & 0.325\\
      & $\gamma_{120}$ & -0.5 & -0.911 & 2.001 & -0.651 & 0.660\\
      & $\gamma_{12Z}$ & -1 & -2.215 & 2.634 & 0.518 & 0.523\\
     \\
      & $\beta_0$ & 1 & 0.006 & 0.053 & -0.001 & 0.052\\
(2)   & $\beta_X$ & -2 & 0.008 & 0.091 & -0.041 & 0.077\\
      & $\gamma_{110}$ & 0.5 & 0.003 & 0.157 & 0.223 & 0.253\\
      & $\gamma_{11Z}$ & 1 & 0.017 & 0.123 & -0.138 & 0.159\\
      & $\gamma_{120}$ & -0.5 & 0.024 & 0.561 & -0.532 & 0.555\\
      & $\gamma_{12Z}$ & -1 & -0.245 & 0.540 & 0.306 & 0.316\\
      \\
      & $\beta_0$ & 1 & -0.008 & 0.095 & -0.102 & 0.125\\
(3)   & $\beta_X$ & -2 & -0.002 & 0.102 & -0.083 & 0.119\\
      & $\gamma_{110}$ & 0.5 & -0.020 & 0.106 & 0.055 & 0.109\\
      & $\gamma_{11Z}$ & 1 & 0.041 & 0.140 & -0.068 & 0.106\\
      & $\gamma_{120}$ & -5 & -0.911 & 1.220 & 3.042 & 3.043\\
      & $\gamma_{12Z}$ & -5 & 0.334 & 0.638 & 4.302 & 4.304\\
     \\
\hline  
\end{tabular}
\begin{tablenotes}
\item[1] All parameters had \textit{Uniform}$(-10, 10)$ prior distributions.
\item[2] All parameters had Normal distributions. Mean and standard deviation specifications are available in Table \ref{sim-setting-table1}.
\end{tablenotes}
\end{threeparttable}
\end{table}

\begin{table}[H]
\centering
\caption{Estimated event probabilities from 500 realizations of simulation Settings 1, 2, and 3. MCMC estimates were computed using the \textit{COMBO} R Package. \cite{COMBO}} \label{appendix-mcmc-prob-table}
\begin{threeparttable}
\begin{tabular}{ccrccccccc}
\hline
\noalign{\vskip 0.2cm}
        Scenario & & & Data\tnote{1} & MCMC & MCMC\\
        
        &&&& Uniform priors\tnote{2}& Normal priors\tnote{3}\\
\hline
    \\
(1)           && $P(Y = 1)$ & 0.647 & 0.652 & 0.625\\
              && $P(Y = 2)$ & 0.353 & 0.348 & 0.375\\
              && $P(Y^* = 1 | Y = 1)$ & 0.847 & 0.874 & 0.874\\
              && $P(Y^* = 2 | Y = 2)$ & 0.877 & 0.940 & 0.862\\
              \\
              \\
(2)           && $P(Y = 1)$ & 0.648 & 0.649 & 0.646 \\
              && $P(Y = 2)$ & 0.352 & 0.351 & 0.354 \\
              && $P(Y^* = 1 | Y = 1)$ & 0.924 & 0.933 & 0.930\\
              && $P(Y^* = 2 | Y = 2)$ & 0.945 & 0.942 & 0.928\\
              \\
              \\
(3)           && $P(Y = 1)$ & 0.647 & 0.646 & 0.630\\
              && $P(Y = 2)$ & 0.353 & 0.354 & 0.370\\
              && $P(Y^* = 1 | Y = 1)$ & 0.846 & 0.859 & 0.855\\
              && $P(Y^* = 2 | Y = 2)$ & 1.000 & 1.000 & 0.946\\
              \\
\hline  
\end{tabular}
\begin{tablenotes}
\item[1] ``Data'' terms refer to empirical values computed from generated datasets.
\item[2] All parameters had \textit{Uniform}$(-10, 10)$ prior distributions.
\item[3] All parameters had Normal distributions. Mean and standard deviation specifications are available in Table \ref{sim-setting-table1}.
\end{tablenotes}
\end{threeparttable}
\end{table}

\end{appendices}

\newpage
\section{Figures}

\newpage

\begin{figure}
\begin{center}
\includegraphics[scale=0.80]{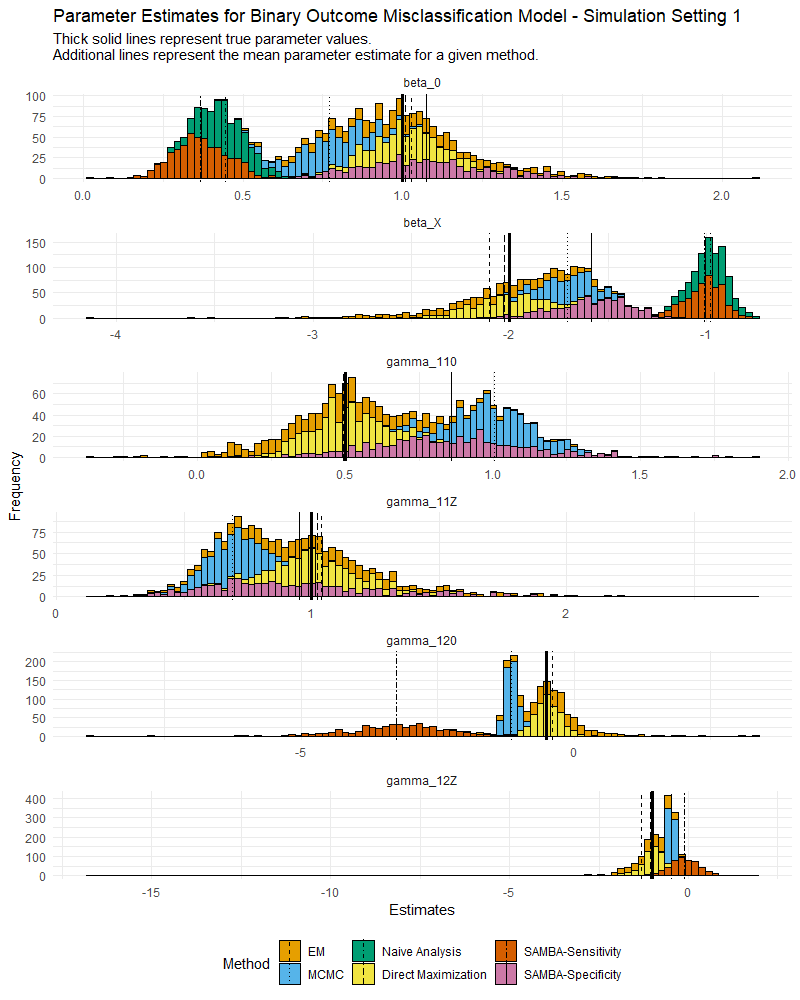}
\medskip
  \caption{Parameter estimates for 500 realizations of simulation Setting 1. All methods are described in Table \ref{parameter-results-table}. Thick solid lines represent true parameter values. Thin lines represent mean parameter estimates for a given method.}\label{fig:small_n_results_figure}
\end{center}
\end{figure}

\begin{figure}
\begin{center}
\includegraphics[scale=0.80]{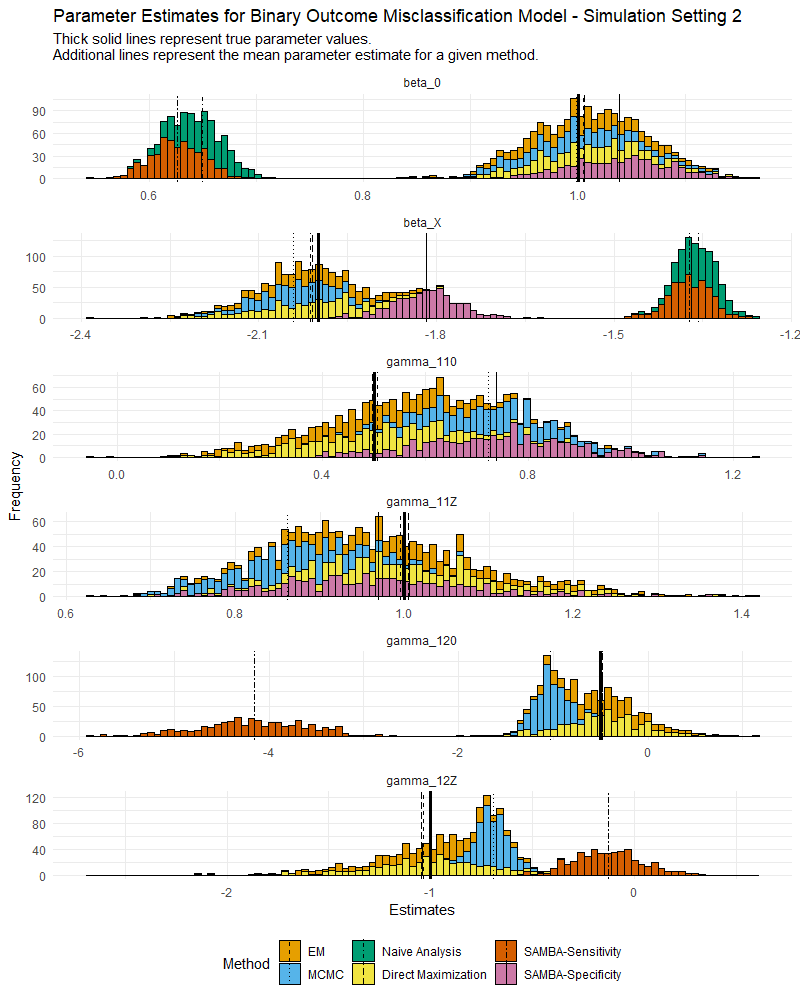}
\medskip
  \caption{Parameter estimates for 500 realizations of simulation Setting 2. All methods are described in Table \ref{parameter-results-table}. Thick solid lines represent true parameter values. Thin lines represent mean parameter estimates for a given method.}\label{fig:large_n_results_figure}
\end{center}
\end{figure}

\clearpage
\section*{Simulation Setting 3 Results Figures} \label{appendix-setting3-figure}

\begin{figure}[h!]
\begin{center}
\includegraphics[scale=.8]{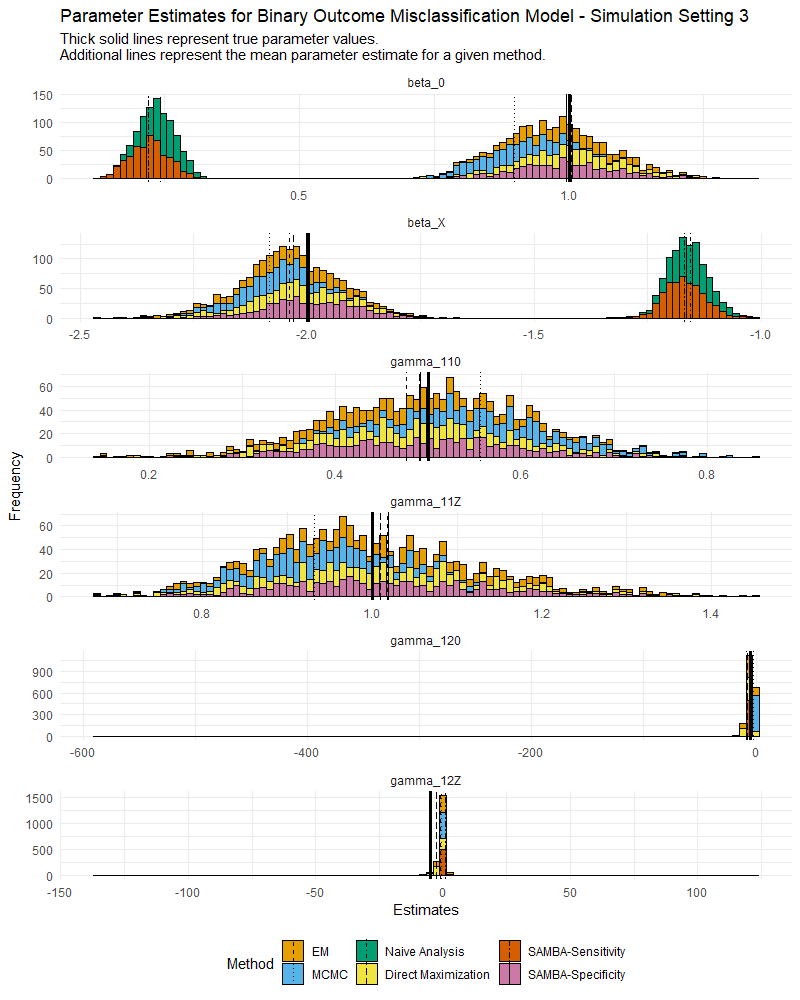}
\medskip
  \caption{Parameter estimates for 500 realizations of simulation Setting 3. All methods are described in Table \ref{parameter-results-table}. Thick solid lines represent true parameter values. Thin lines represent mean parameter estimates for a given method.}\label{fig:ps_sim_results_histogram}
\end{center}
\end{figure}

\begin{figure}[h!]
\begin{center}
\includegraphics[scale=.85]{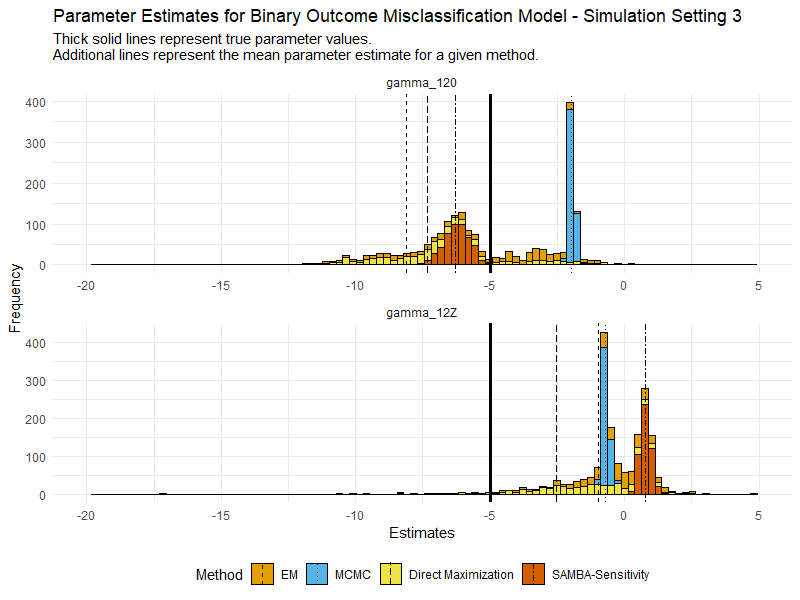}
\medskip
  \caption{$\gamma_{120}$ and $\gamma_{12Z}$ parameter estimates for 500 realizations of simulation Setting 3. All methods are described in Table \ref{parameter-results-table}. Thick solid lines represent true parameter values. Thin lines represent mean parameter estimates for a given method. The x-axis of this plot is truncated between $-20$ and $5$ to enable comparison between methods within that range. }\label{fig:ps_sim_results_histogram_zoom}
\end{center}
\end{figure}

\end{document}